\documentclass[aps,prd,preprint,groupedaddress,amssymb,amsmath,nofootinbib]{revtex4}

\usepackage{color}
\usepackage{graphicx}
\usepackage{epstopdf}
\usepackage{slashed}

\usepackage{epsf,epsfig,subfigure,graphicx,amsmath,amssymb,mathtools}

\usepackage{hhline}

\begin{document}

\preprint{}

\title{Effects of vectorlike leptons on $h\to 4\ell$ \\
and the connection to the muon $g-2$ anomaly.}

\author{Radovan Derm\' \i\v sek}
\email[]{dermisek@indiana.edu}
\affiliation{Physics Department, Indiana University, Bloomington, IN 47405, USA}

\author{Aditi Raval}
\email[]{adiraval@indiana.edu}
\affiliation{Physics Department, Indiana University, Bloomington, IN 47405, USA}

\author{Seodong Shin}
\email[]{shinseod@indiana.edu}
\affiliation{Physics Department, Indiana University, Bloomington, IN 47405, USA}


\date{June 25, 2014}

\begin{abstract}
The mixing of new vectorlike leptons with leptons in the standard model can generate  flavor violating couplings of $h$, $W$ and $Z$ between heavy and light leptons. Focusing on the couplings of the muon,  the partial decay width of  $h\to e_4^\pm \mu^\mp$, where $e_4$ is the new lepton, can be significant  when this process is kinematically allowed. Subsequent decays $e_4^\pm \to Z\mu^\pm$ and $e_4^\pm \to W^\pm \nu$  lead to the same final states as $h \to ZZ^* \to Z \mu^+\mu^-$ and    $h \to WW^* \to W \mu\nu$, thus possibly affecting measurements of these processes. We calculate $h\to e_4 \ell_i
 \to Z\ell_i\ell_j$, where $\ell_{i,j}$ are standard model leptons, including the possibility of off-shell decays, interference with $h\to ZZ^* \to Z \ell_i \ell_i$, and the mass effect of $\ell_{i,j}$ which are important when the mass of $e_4$ is close to the mass of the Higgs boson. We derive constraints on masses and couplings of the heavy lepton from the measurement of $h\to 4\ell$. We focus on the couplings of the muon and discuss possible effects on $h\to ZZ^*$ from the region of parameters that can explain the anomaly in the measurement of the muon $g-2$.

\end{abstract}


\maketitle

 \section{Introduction}
  
  Among simple extensions of the standard model (SM) are those with extra vectorlike fermions near the electroweak scale. Vectorlike fermions can acquire masses independently  of their Yukawa couplings to the Higgs boson and thus are not strongly constrained (compared to chiral fermions) by experiments. However, even small Yukawa couplings between  SM fermions and vectorlike fermions can affect a variety of processes, including  Higgs boson decays.
  
We consider the extension of the SM with extra SU(2) doublet, $L_L$, and singlet, $E_R$, leptons (with the same hypercharges as SM leptons) and  their vectorlike partners.    
The mixing of new vectorlike leptons with leptons in the SM can generate  flavor violating couplings of $h$, $W$ and $Z$ between heavy and light leptons. Focusing on the couplings of the muon,  the partial decay width of  $h \to e_4^\pm \mu^\mp$, where $e_4$ is the lightest new mass eigenstate, can be significant if this process is kinematically allowed. Subsequent decays of the heavy lepton, $e_4^\pm \to Z\mu^\pm$ and $e_4^\pm \to W^\pm \nu$,  lead to the same final states as $h \to ZZ^* \to Z \mu^+\mu^-$ and    $h \to WW^* \to W \mu\nu$, thus possibly affecting measurements of these processes. Since the partial width of $h \to Z\mu^+\mu^-$ is much smaller than $h \to W \mu\nu$ in the SM,  we mainly focus on $e_4^\pm \to Z\mu^\pm$. 

We calculate $h\to e_4 \ell_i
 \to Z\ell_i\ell_j$, where $\ell_{i,j}$ are SM leptons, including the possibility of off-shell decays, interference with $h\to ZZ^* \to Z \ell_i \ell_i$, and the mass effect of $\ell_{i,j}$ which are important when the mass of $e_4$ is close to the mass of the Higgs boson. We derive constraints on masses and couplings of the heavy lepton from the searches for $h\to ZZ^*\to4\ell$ at ATLAS~\cite{atlas4l} and CMS~\cite{cms4l}. 
 
 Although our calculation is general and can be presented for any final state, we focus on the couplings of the muon and discuss possible effects on $h\to ZZ^* \to 4\mu$ or $2e2\mu$  from the region of parameters that can explain the anomaly in the measurement of the muon $g-2$.
  It has been shown that the discrepancy between the measured value of the muon anomalous magnetic moment and the SM prediction can be explained by contributions from extra vectorlike leptons~\cite{Kannike:2011ng, Dermisek:2013gta}. A particularly interesting solution to the muon $g-2$ simultaneously explaining the muon mass completely from the mixing of the muon with vectorlike leptons requires the mass of the lepton doublet to be within about 130 GeV~\cite{Dermisek:2013gta}. Thus in a large range of the parameter space this solution predicts the existence of $e_4$ below the Higgs mass and thus $h \to e_4^\pm \mu^\mp$ could be kinematically open and potentially significant. The $e_4 - \mu - h $, $e_4 - \nu - W $ and $e_4 - \mu - Z $ couplings needed to explain the muon $g-2$ anomaly are sufficient to modify the Higgs decays in $4\ell$ and $2\ell 2\nu$ channels. Thus the contributions to the muon $g-2$ and $h\to 4\ell$ can be connected without any further assumptions. The correlation with contributions to other Higgs decays, $h\to \mu^+ \mu^-$ and $h \to \gamma  \gamma$,  can be also found in~\cite{Dermisek:2013gta}.

Flavor violating Higgs decays into pairs of SM fermions were previously studied in Refs.~\cite{Blankenburg:2012ex,  Harnik:2012pb}. These can also  be induced by mixing of SM fermions with vectorlike fermions; however, we do not consider this possibility here.   
We only allow one of the SM leptons to mix with  vectorlike leptons in which case the flavor violating couplings to SM leptons are not generated.

In general, vectorlike quarks and leptons near the electroweak scale provide very rich phenomenology. For a recent discussion see for example Ref.~\cite{Ellis:2014dza} and references within. The addition of three or more  complete vectorlike families also provides a simple UV completion of the SM featuring gauge coupling unification, sufficiently stable proton, and the Higgs quartic coupling remaining positive all the way to the grand unified theory (GUT) scale~\cite{Dermisek:2012as, Dermisek:2012ke}.

   This paper is organized as follows. In Sec.~\ref{sec:model}, we briefly summarize the general framework and discuss constraints on possible flavor violating couplings between the muon and a heavy lepton.
 In Sec.~\ref{sec:Zmumu}, we calculate the effect of  $h \to e_4^\pm \mu^\mp \to Z \mu^+ \mu^-$ on $h\to 4\ell$. We also discuss a connection with the explanation of the muon $g-2$ anomaly in Sec.~\ref{sec:g-2} and 
  provide some concluding remarks in Sec.~\ref{sec:conclusions}. In Appendixes we extract bounds on $h\to 4\mu$ and $h \to 2e2\mu$ from ATLAS and CMS searches, compare the ATLAS and CMS  limits on couplings and masses of the new lepton, and calculate the partial width of a scalar to four leptons in the presence of general flavor violating couplings of the new lepton. These formulas include the mass effects of final state leptons and interference with $h\to ZZ^*$.  We also briefly comment on the impact of constraints from $h \to WW^*\to 2\ell 2\nu$  which are typically weaker than those from $h \to ZZ^*\to 4\ell$ , unless BR($e_4^\pm \to W^\pm \nu$) is close to 1.

 \section{Outline of the Framework}
 \label{sec:model}
 
 We extend the SM by vectorlike pairs of new leptons, $L_{L,R}$ and $E_{L,R}$, where $E_R$ ($L_{L}$) has the same quantum numbers as $\mu_R$ ($\mu_L$) in the SM, and $E_L$ ($L_{R}$) is its vectorlike partner. For SU(2) doublets we use the same label for their charged components as for the whole doublets. We assume that the new leptons mix only with one SM  lepton and we take the muon as an example.  The results for the electron or tau lepton  could be obtained in the same way. If the new leptons mix simultaneously with more than one SM lepton, the generated flavor violating couplings need to satisfy all the constraints from a variety of  processes involving SM fermions. We will not pursue this direction here, and for simplicity we assume that all other couplings are zero.

With this assumption, the most general renormalizable Lagrangian for the muon and new leptons is given by:
\begin{eqnarray}
{\cal L} &\supset& -  \bar \mu_{L} y_{\mu} \mu_{R} H - \bar \mu_{L} \lambda^E E_{R} H   - \bar L_{L} \lambda^L \mu_{R} H -  \lambda \bar L_{L}  E_{R} H - \bar \lambda H^\dagger \bar E_{L}  L_{R}  \nonumber \\ 
&& - M_L \bar L_L L_R - M_E \bar E_L E_R + {\it h.c.},
\label{eq:lagrangian}
\end{eqnarray}
where the first term is the usual SM Yukawa coupling,  followed by Yukawa couplings between the muon and vectorlike leptons, Yukawa couplings between  vectorlike leptons, and  mass terms for vectorlike  leptons.

After spontaneous symmetry breaking, $H = ( 0, v + h/\sqrt{2} )^T$,
the resulting mass matrix  for the muon and the extra leptons can be diagonalized by a biunitary transformation: 
\begin{eqnarray}
U^\dagger_L
\begin{pmatrix}
 y_\mu v & 0 &   \lambda^E v\\
  \lambda^L v & M_L &  \lambda v\\
 0 & \bar \lambda v & M_E \\
\end{pmatrix}
U_R 
 =  
 \begin{pmatrix}
m_\mu  & 0 &   0\\
 0 & m_{e_4} &  0\\
 0 & 0 & m_{e_5} \\
\end{pmatrix},
\label{eq:mass}
\end{eqnarray}
where we label the new mass eigenstates by $e_4$ and $e_5$.

Couplings of  the muon and heavy leptons  to the $Z$, $W$ and Higgs bosons are modified  because the  $E_L$  is an SU(2) singlet mixing with other SU(2) doublets, and the charged component of $L_R$ which originates from an SU(2) doublet mixes with other SU(2) singlets. The flavor conserving couplings receive corrections and flavor violating couplings between the muon and heavy leptons are generated. These couplings are given in Ref.~\cite{Dermisek:2013gta} in terms of diagonalization matrices. The forms of diagonalization matrices $U_{L,R}$, which are useful for deriving approximate formulas for  couplings of $Z, W$ and $h$, are also given in Ref.~\cite{Dermisek:2013gta}  in the limit $\lambda^E v, \lambda^L v, \bar \lambda v, \lambda v  \ll M_E, M_L$. 

In what follows we assume that only one new lepton is below or close to the Higgs mass and we define the couplings of the lighter new  lepton, $e_4$, and the muon to the $Z$ and Higgs bosons  by the effective Lagrangian of the form
\begin{equation}
{\cal L} \supset   \;   g^{Z}_L  \, \bar e_{4L} \gamma^\mu  \mu_{L} Z_\mu  \; +\;  g^{Z}_R \, \bar e_{4R} \gamma^\mu   \mu_{R} Z_\mu 
  \; -\;  \frac{1}{\sqrt{2}} g^{h}_R \, \bar e_{4L}   \mu_{R} h \; -\;  \frac{1}{\sqrt{2}} g^{h}_L \, \bar e_{4R}   \mu_{L} h \; + h.c..
  \label{eq:eff_lagrangian}
\end{equation}
The formulas for these couplings, and all other couplings (couplings to $W$ and flavor conserving couplings) in terms of Lagrangian parameters can be found in Ref.~\cite{Dermisek:2013gta}.

In order to satisfy precision electroweak data
 related to the muon that include the $Z$ pole observables ($Z$ partial width, forward-backward asymmetry, left-right asymmetry), the $W$  partial width, the muon lifetime and constraints from oblique corrections, namely from $S$ and $T$ parameters,
possible values of $g^{Z}_L$ and $g^{Z}_R$  are constrained to be smaller than about $0.01$ and $0.015$. The maximum allowed values of the Higgs couplings depend on the size  of the Yukawa couplings in Eq.~(\ref{eq:lagrangian}). Limiting all couplings to be smaller than 0.5 (1.0) the $g^{h}_L$ coupling is limited to 0.03 (0.06) and the $g^{h}_R$  coupling is limited to 0.04 (0.08). We also impose the LEP limit, 105 GeV, on the mass of the new charged lepton.

\section{$h \to e_4^\pm \mu^\mp \to Z \mu^+ \mu^-$ and $h \to ZZ^* \to Z\mu^+\mu^-$}
\label{sec:Zmumu}
   
\begin{figure}[t,h]
\includegraphics[width=2.in]{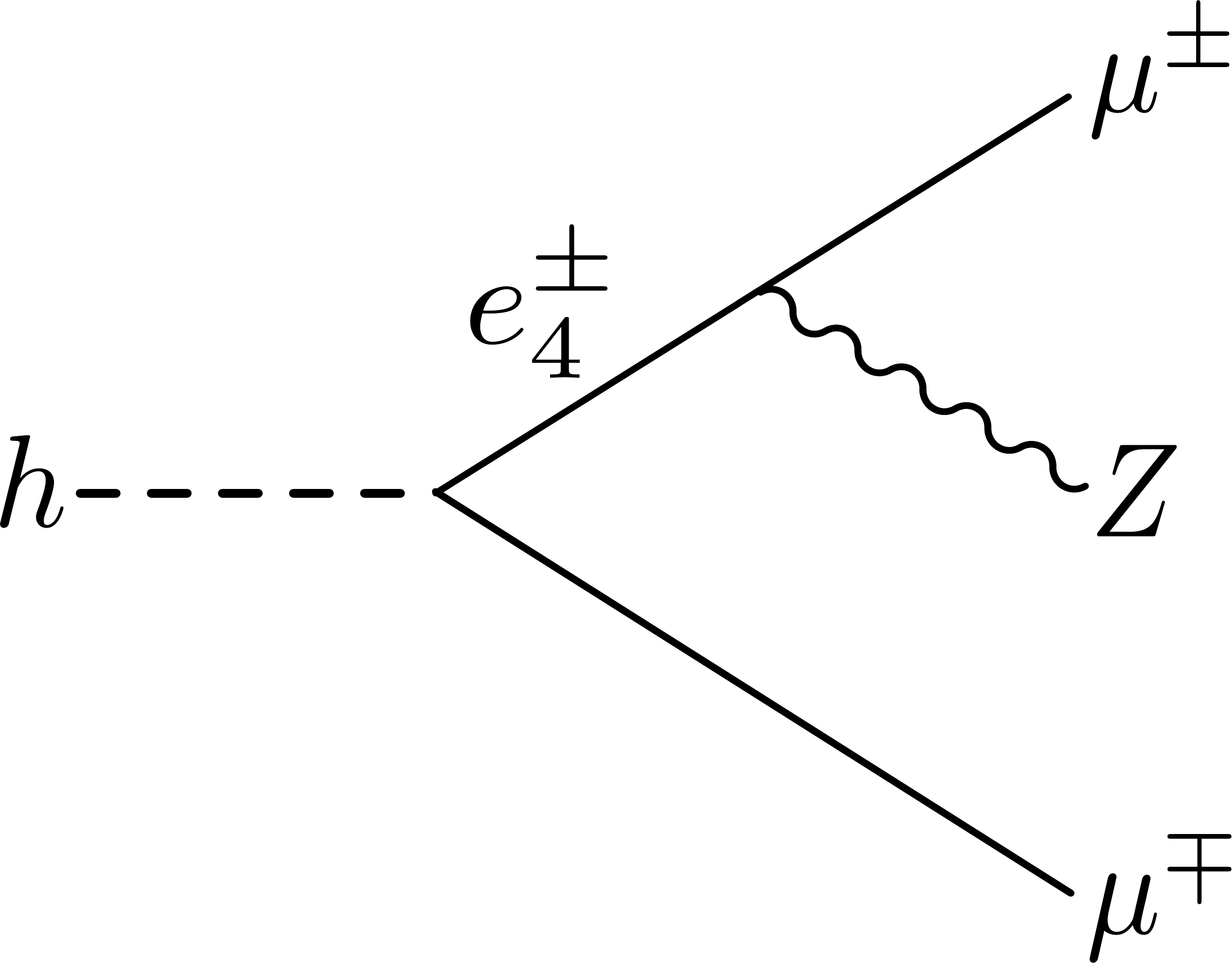} 
\caption{The Feynman diagram for $h \to e_4^\pm \mu^\mp \to Z \mu^+ \mu^-$ contributing to the same final state as $h\to ZZ^* \to Z \mu^+ \mu^-$.}
\label{fig:htoZmumu}
\end{figure}

Even a small  flavor violating coupling of the Higgs boson to a new charged lepton and the muon can lead to a sizable contribution to the Higgs width if $h \to e_4^\pm \mu^\mp$ is kinematically open. The decay mode of the new lepton, $e_4^\pm \to Z \mu^\pm$, leads to $h \to Z \mu^+ \mu^-$ final state (see Fig.~\ref{fig:htoZmumu}),  which is the same as the final state of $h \to ZZ^* \to Z \mu^+ \mu^-$. Thus the new charged lepton can contribute to the $h\to 4\mu$ and $h\to 2e 2\mu$ processes. Without additional couplings it cannot contribute to $h\to 2\mu 2e$ (the first pair of leptons originating from the on-shell $Z$) or $h\to 4e$ decay modes.
   
The approximate formula for the partial decay width of $h \to e_4^\pm \mu^\mp$, neglecting the mass of the muon, is given by
 \begin{equation}
\Gamma(h \to e_4^\pm \mu^\mp) \simeq \frac{m_h}{16\pi}  \left[ (g_L^h)^2 + (g_R^h)^2 \right] \left( 1 - \frac{m_{e_4}^2 }{m_h^2} \right)^2
\end{equation}
and the  formula for $e_4^\pm \to Z \mu^\pm$ in the same approximation is given by
\begin{equation}
\Gamma(e_4^\pm \to Z \mu^\pm) = \frac{m_{e_4}}{32 \pi}  \left[ (g_L^Z)^2 + (g_R^Z)^2 \right] \frac{m_{e_4}^2}{M_Z^2} \left( 1 - \frac{M_Z^2}{m_{e_4}^2} \right)^2 \left( 1 + 2\frac{M_Z^2}{m_{e_4}^2} \right). 
\end{equation}
The complete formula for $h \to (e_4^\pm\mu^\mp,ZZ^*)\to Z\mu^+\mu^-$ and actually for any fermions in the final state, including the mass effect of final state fermions, the contribution from off-shell $e_4$ and the  interference with $h\to ZZ^*$ is given in Appendix~\ref{app:width}.

Although the $4\ell$ final states originating from $h \to e_4^\pm \mu^\mp$ and  $h \to ZZ^*$ are identical, the kinematical distribution of final state leptons is not. The muon that accompanies the $e_4$ is somewhat soft, and if the mass of $e_4$ is close to the Higgs mass, this muon does not pass the cuts used in the $h\to ZZ^*$ analysis. To quantify the contribution of the   $h \to e_4^\pm \mu^\mp$ to  $h \to Z\mu^+\mu^-$ we define
 \begin{equation}
R_{Z\mu\mu} = \xi \frac{ \Gamma(h \to (e_4^\pm\mu^\mp,ZZ^*)\to Z\mu^+\mu^-)}{\Gamma(h \to ZZ^*\to Z\mu^+\mu^-)_{SM}}  ,
\label{eq:R_Zmumu}
\end{equation}
 where $\xi $ is the  acceptance of the SM+new lepton contribution to $h\to Z\mu^+\mu^-$ relative to the SM $h\to ZZ^* \to Z \mu^+\mu^-$. Because of the interference the contributions from new physics and  the SM do not factor out.

\begin{figure}[t,h]
\includegraphics[width=3.in]{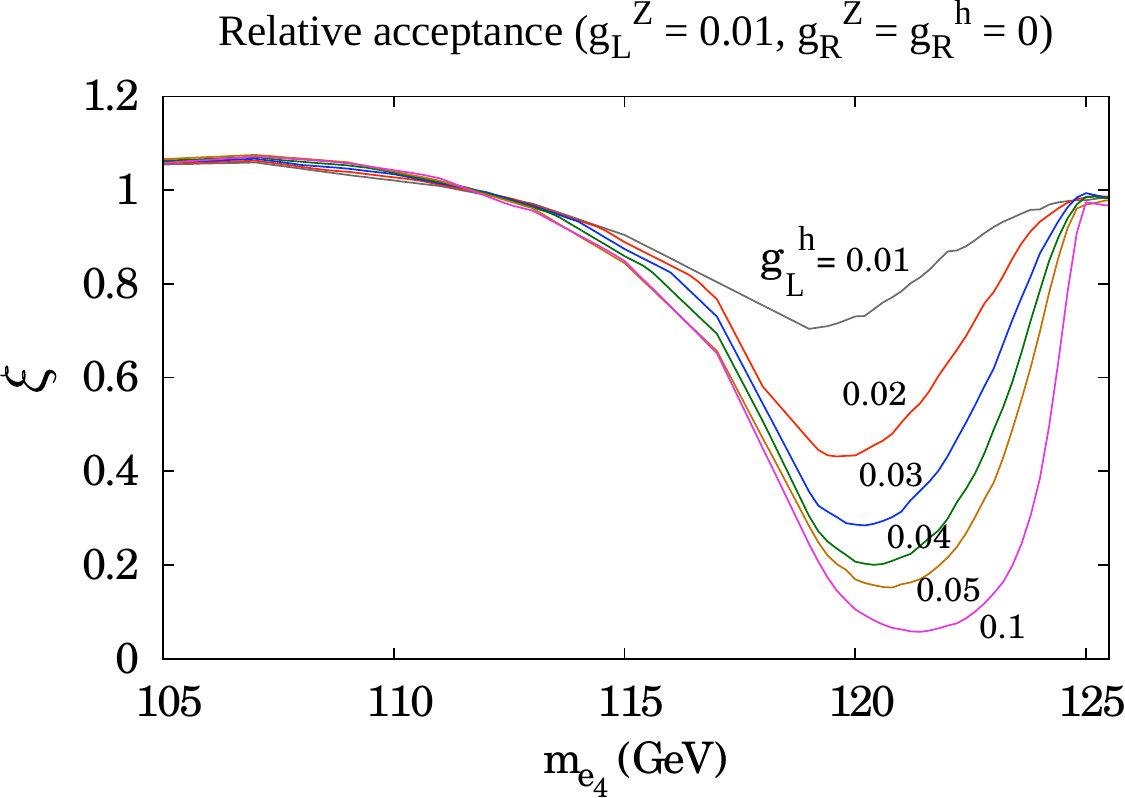} 
\caption{The Atlas acceptance of $h \to (e_4^\pm\mu^\mp,ZZ^*)\to 4\mu$ relative to the $h \to ZZ^*\to 4\mu$ in the SM.}
\label{fig:acceptance}
\end{figure}

 The relative acceptance for $h\to 4\mu$ case is given in Fig.~\ref{fig:acceptance} as a function of the mass of the $e_4$ for various values of $g_L^h$ coupling. We adopt  the cuts from the ATLAS analysis~\cite{atlas4l}. For sizable $g_L^h$ coupling the $h \to e_4^\pm \mu^\mp$ would easily dominate over $h \to Z\mu^+\mu^-$.  As the mass of the $e_4$ increases, the accompanying muon is getting softer and eventually does not pass the cuts in the $h\to ZZ^*$ analysis. Thus, the acceptance is dropping significantly at about 6 GeV from the Higgs mass. 
 Close to the kinematical threshold it increases again since the SM contribution dominates.
 As the $g_L^h$ coupling decreases, the SM contribution starts to dominate and the relative acceptance is getting close to 1. 
 The $g_L^Z$  coupling is set to 0.01 and other couplings are set to zero in Fig.~\ref{fig:acceptance}. The results do not depend much on the choice of $g_L^Z$ since it only enters through the interference with the SM contribution or when the $e_4$ is not on shell. The results for other coupling combinations are almost identical since the interference is small (the interference changes sign when $g_L^Z \leftrightarrow g_R^Z$).

\begin{figure}[t]
 \includegraphics[width=3.in]{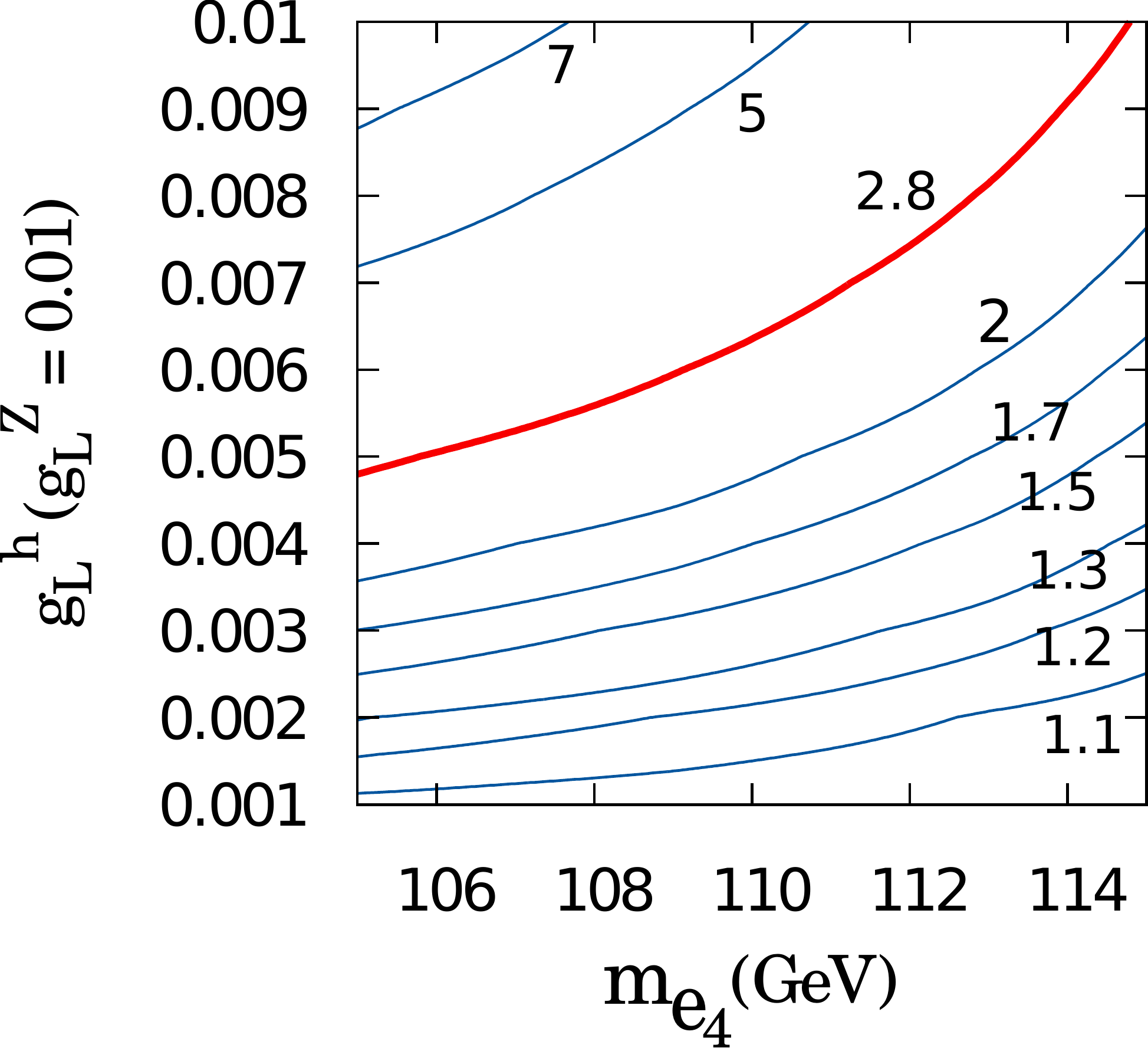} \hspace{0.5cm}
\includegraphics[width=2.9in]{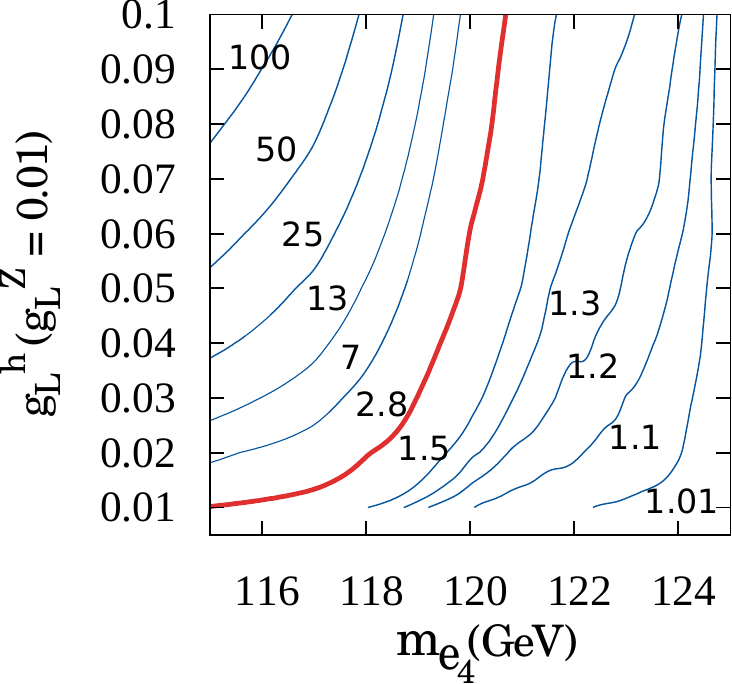}
\caption{Contours of constant $R_{Z\mu\mu}$ in $g_L^h - m_{e_4}$ plane for fixed $g_L^Z = 0.01$ and $g_R^h = g_R^Z = 0$ assuming BR$(e_4 \to Z \mu) = 100\%$. The thick line represents the ATLAS exclusion limit for $4\mu$ final state. 
For smaller BR$(e_4 \to Z \mu)$ predicted values of $R_{Z\mu\mu}$ and the exclusion limit can be obtained by simple rescaling.
 }
\label{fig:R_Zmumu}
\end{figure}

 Predictions for $R_{Z\mu\mu}$ as a function of $g_L^h $ and $m_{e_4}$ for fixed $g_L^Z = 0.01$ and $g_R^h = g_R^Z = 0$ are given in Fig.~\ref{fig:R_Zmumu}.  The thick line represents the ATLAS upper exclusion limit for $4\mu$ final state, which is discussed in detail in  Appendix~\ref{app:limits}.  The plots assume 
 BR$(e_4 \to Z \mu) = 100\%$.  For smaller BR$(e_4 \to Z \mu)$ predicted values of $R_{Z\mu\mu}$ and the exclusion limit can be obtained by simple rescaling. As is easily seen  a large region of possible couplings  and  masses of the $e_4$ is excluded. However, for $m_{e_4} \gtrsim 120$ GeV  the $h\to 4\mu$ does not exclude any scenario that would not be already excluded by precision electroweak (EW) data.

 Defining the average Higgs coupling,
   \begin{equation}
 g_h \equiv \sqrt{(g_L^h)^2+(g_R^h)^2},
 \label{eq:gh}
 \end{equation}
  which approximately controls the partial width of $h \to e_4^\pm \mu^\mp$, 
 the $y$-axis of the plots in Fig.~\ref{fig:R_Zmumu} could be very well approximated by $g_h \sqrt{{\rm BR}(e_4\to Z\mu)}$ when $e_4$ is on-shell.

 \section{Connection with the muon $g-2$ anomaly}
 \label{sec:g-2}

 The discrepancy between the measured value of the muon anomalous magnetic moment~\cite{Bennett:2006fi} and the SM prediction, $
 \Delta a^{exp}_\mu	= a^{exp}_\mu - a^{SM}_\mu = 2.7 \pm 0.80 \times 10^{-9},$ is at the level of 3.4 standard deviations. It can be explained by contributions from extra charged lepton (originating either from $L$ and $E$) in the loop diagram with the Higgs and $Z$ bosons, and extra neutrino (originating from $L$) in the loop diagram with the $W$ boson~\cite{Kannike:2011ng}\cite{Dermisek:2013gta}.

In Ref.~\cite{Dermisek:2013gta} it was shown that there are two generic solutions to the muon $g-2$ that differ in the correlation between the contribution of the vectorlike leptons to the muon mass, $ m_\mu^{LE}$,  and the muon $g-2$.
 This correlation is controlled by $M_L$ which represents the mass of extra neutrino $\nu_4$. In the asymptotic solution,  $M_L \gg M_Z$,  the Higgs loop dominates and the measured value of the muon $g-2$ is obtained for $ m_\mu^{LE}/m_\mu \simeq -1$. In this case, the physical muon mass is a result of a cancellation between twice as large  direct Yukawa coupling and the contribution from the mixing with heavy leptons. The second solution is with a light extra neutrino, $M_L \simeq M_Z$, in which case the $W$ loop dominates and the measured value of the muon $g-2$ is obtained for $ m_\mu^{LE}/m_\mu \simeq +1$. In this case, the muon mass can fully originate from the mixing with heavy leptons.

The sizes of possible contributions from vectorlike leptons to various observables depend on the upper limit on Yukawa couplings that we allow in the model. The upper limit 0.5 is sufficient to fully explain the muon $g-2$ anomaly and generate the muon mass and is small enough so that the model can be embedded into a scenario with three complete vectorlike families which provides a simple UV embedding (with gauge coupling unification, sufficiently stable proton, and the Higgs quartic coupling remaining positive all the way to the GUT scale)~\cite{Dermisek:2012as, Dermisek:2012ke}. With this upper limit, the muon $g-2$ can be explained within one standard deviation  with  $M_L \lesssim 130$ GeV and the mass of the lightest extra charged lepton originating mostly from $L$, $m_{e_4} \lesssim 150$ GeV. The mass of $e_4$ is not strictly  given by $M_L$ due to possible mixing. In about half of the allowed region  the mass of $e_4$ is below the Higgs boson mass and thus $h \to e_4^\pm \mu^\mp$ can significantly contribute  to $h \to 4\mu$ $ (2e2\mu)$. The asymptotic solution requires, $M_L \gtrsim 1$ TeV, but even in this case, the other charged lepton, originating mostly from $E$,  can be  below the Higgs mass. We will present results for both cases.

In this section, we use the same points in the parameter space generated for Ref.~\cite{Dermisek:2013gta} so that the correlations with other observables studied in~\cite{Dermisek:2013gta}, $h \to \mu^+ \mu^-$ or $h \to \gamma \gamma$, can be easily inferred. However, we impose new  constraints on $h \to \mu^+ \mu^-$  from CMS, which limit
  \begin{equation}
 R_{\mu\mu} \equiv \frac{\Gamma(h \to \mu^+ \mu^-)}{\Gamma(h \to \mu^+ \mu^-)_{SM}}.
 \label{eq:Rmumu}
 \end{equation}
to $R_{\mu\mu}^{exp} \leq 7.4$~\cite{CMS_h_to_mumu}. It is interesting to note that there is a slight excess of events for the reconstructed dimuon mass near the measured Higgs mass. The small $M_L$ solution to the muon $g-2$ predicts enhancement of the  $h \to \mu^+ \mu^-$ by a factor of 5 -- 9.

  \begin{figure}[t]
\includegraphics[width=3.in]{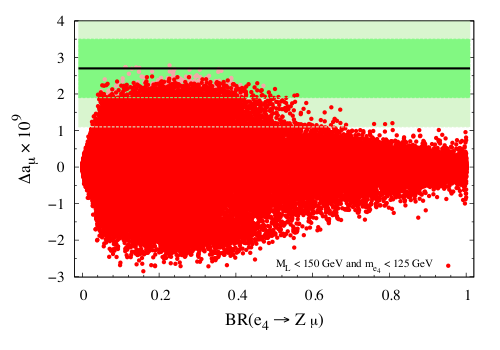} \hspace{0.5cm}
\includegraphics[width=3.in]{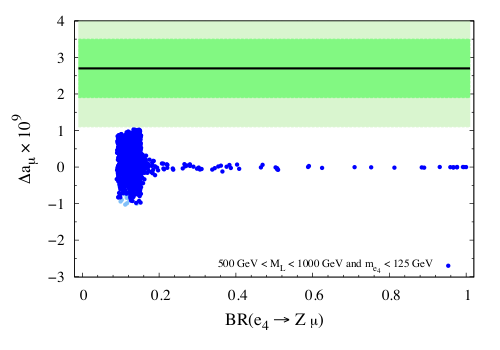}
\caption{Randomly generated points with $M_L \in [100, 150]$ GeV (left) and $M_L \in [500, 1000]$ GeV (right), with $M_E \in [100, 1000]$ GeV,  $\lambda, \; \bar \lambda < 0.5$, and $\lambda^{L,E}$  in allowed ranges from precision EW data, plotted in $\Delta a_\mu$ -- BR$(e_4\to Z\mu)$ plane. The lightest mass eigenstate is required to satisfy the LEP limit. This is a subset of the points  generated in Ref.~\cite{Dermisek:2013gta}  which have one mass eigenstate below the Higgs mass 125 GeV. The lightly shaded points are excluded by the CMS search for $h\to \mu^+\mu^-$.
The horizontal line and dark (light) shaded bands correspond to the central experimental value of $\Delta a_\mu$ and $1\sigma$ ($2\sigma$) regions, respectively.
 }
\label{fig:p5p5}
\end{figure}

 \begin{figure}[t]
\includegraphics[width=3.in]{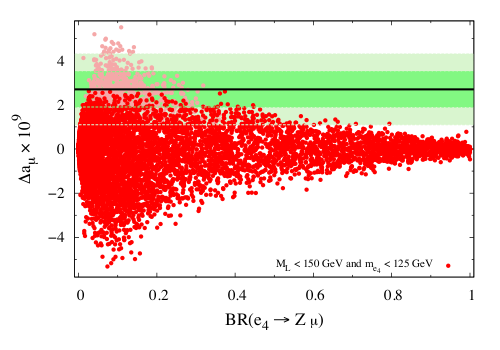} \hspace{0.5cm}
\includegraphics[width=3.in]{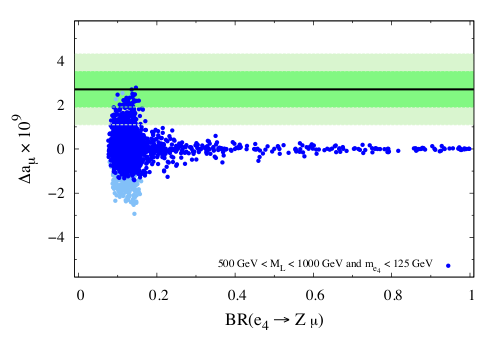}
\caption{The same as in Fig.~\ref{fig:p5p5} with ranges of Yukawa couplings $\lambda, \; \bar \lambda$ extended to  1.
 }
\label{fig:1p1p}
\end{figure}

In Figs.~\ref{fig:p5p5} and \ref{fig:1p1p} we plot the contribution to the muon $g-2$ versus BR$(e_4\to Z\mu)$ for the subset of the points generated in Ref.~\cite{Dermisek:2013gta}  which have one mass eigenstate below the Higgs mass, 125 GeV. Points with $M_L \in [100, 150]$ GeV are plotted on the left and those with $M_L \in [500, 1000]$ GeV are on the right. The lightly shaded points show the impact of  the CMS search for $h\to \mu^+\mu^-$. In Fig.~\ref{fig:p5p5} the Yukawa couplings in Eq.~(\ref{eq:lagrangian}) are limited to 0.5, while in Fig.~\ref{fig:1p1p} their possible values are extended to 1.  Precision EW data and  the LEP limit on a new charged lepton mass, 105 GeV, are satisfied by all points.
We see that the points that can explain the muon $g-2$ within 1 standard deviation predict BR$(e_4\to Z\mu)$ between a few and 50\% for the small $M_L$ solution, and about 10\% for the asymptotic case. The rest of the width of $e_4$ is given by $e_4\to W\nu$, since $e_4$ cannot decay to the Higgs boson.

\begin{figure}[t]
\includegraphics[width=3.in]{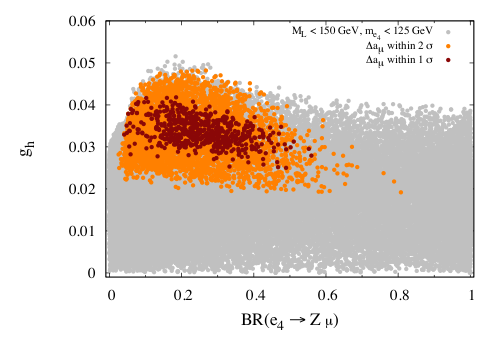} \hspace{0.5cm}
\includegraphics[width=3.in]{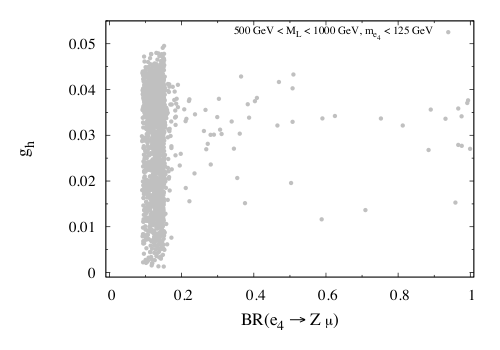}
\caption{Points from Fig.~\ref{fig:p5p5} that satisfy the CMS limit on $h\to \mu^+\mu^-$ plotted in $g_h$ -- BR$(e_4\to Z\mu)$ plane. The darker points (orange points on the left) 
satisfy the muon $g-2$ within $2\sigma$ while the darkest points (dark red  points on the left)
satisfy the muon $g-2$ within $1\sigma$.   }
\label{fig:p5p5_gh}
\end{figure}

\begin{figure}[t]
\includegraphics[width=3.in]{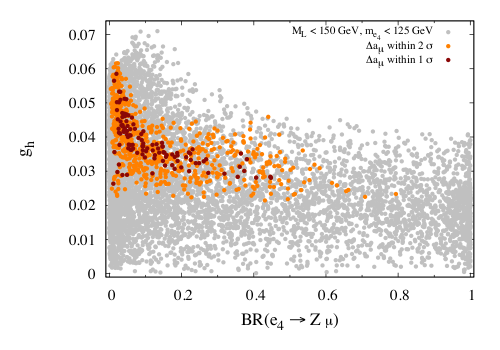} \hspace{0.5cm}
\includegraphics[width=3.in]{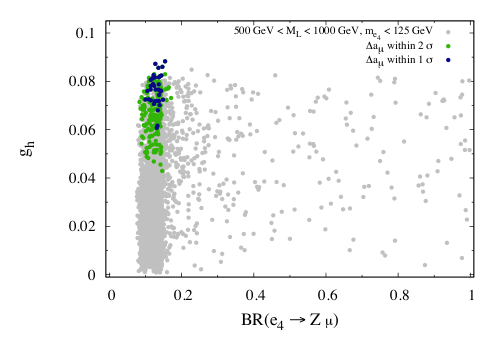}
\caption{Points from Fig.~\ref{fig:1p1p}  that satisfy the CMS limit on $h\to \mu^+\mu^-$ plotted in $g_h$ -- BR$(e_4\to Z\mu)$ plane.  The darker points (orange points on the left and green  points on the right) satisfy the muon $g-2$ within $2\sigma$ while the darkest points (dark red  points on the left and dark blue points on the right) satisfy the muon $g-2$ within $1\sigma$.}
\label{fig:1p1p_gh}
\end{figure}

The correlation between BR$(e_4\to Z\mu)$  and the average Higgs coupling $g_h$ defined in Eq.~(\ref{eq:gh}) is shown in Figs.~\ref{fig:p5p5_gh} and \ref{fig:1p1p_gh}. The plotted points are all the points from Figs.~\ref{fig:p5p5} and \ref{fig:1p1p} that satisfy the CMS limit on $h\to \mu^+\mu^-$. The darker points (orange points on the left and green  points on the right) satisfy the muon $g-2$ within $2\sigma$ while the darkest points (dark red  points on the left and dark blue points on the right) satisfy the muon $g-2$ within $1\sigma$.

\begin{figure}[t]
\includegraphics[width=3.in]{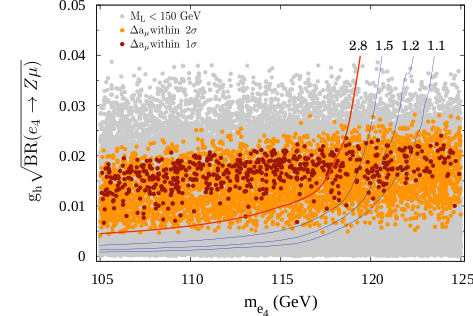}
\caption{Points from Fig.~\ref{fig:p5p5} (left) that satisfy the CMS limit on $h\to \mu^+\mu^-$ plotted in $g_h \sqrt{{\rm BR}(e_4\to Z\mu)}$ -- $m_{e_4}$ plane. The darker (orange) points  satisfy the muon $g-2$ within $2\sigma$ while the darkest (dark red) points satisfy the muon $g-2$ within $1\sigma$.  Overlaid are selected contours of constant $R_{Z\mu\mu}$ from Fig.~\ref{fig:R_Zmumu}. }
\label{fig:p5p5_gh_me4}
\end{figure}

\begin{figure}[t]
\includegraphics[width=3.in]{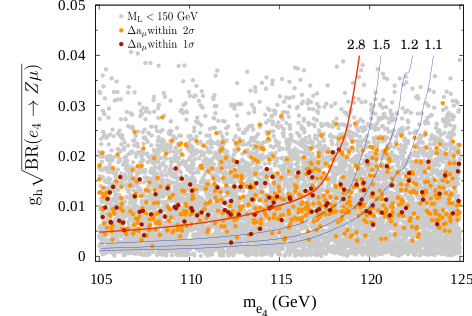} \hspace{0.5cm}
\includegraphics[width=2.8in]{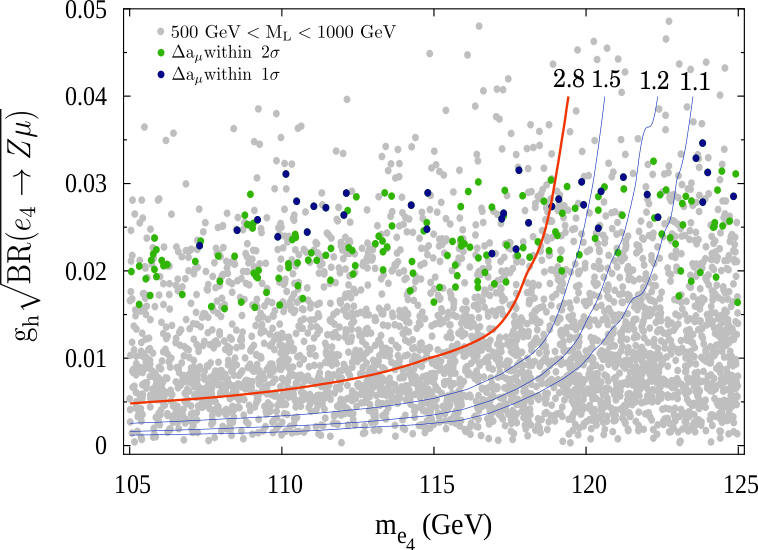}
\caption{Points from Fig.~\ref{fig:1p1p}  that satisfy the CMS limit on $h\to \mu^+\mu^-$ plotted in $g_h \sqrt{{\rm BR}(e_4\to Z\mu)}$ -- $m_{e_4}$ plane. The darker points (orange points on the left and green  points on the right) satisfy the muon $g-2$ within $2\sigma$ while the darkest points (dark red  points on the left and dark blue points on the right) satisfy the muon $g-2$ within $1\sigma$. Overlaid are selected contours of constant $R_{Z\mu\mu}$ from Fig.~\ref{fig:R_Zmumu}.
 }
\label{fig:1p1p_gh_me4}
\end{figure}

 Finally, in Figs.~\ref{fig:p5p5_gh_me4} and \ref{fig:1p1p_gh_me4} we show the same points  in $g_h \sqrt{{\rm BR}(e_4\to Z\mu)}$ -- $m_{e_4}$ plane. The combination $g_h \sqrt{{\rm BR}(e_4\to Z\mu)}$ very well approximates the $y$ axis of Fig.~\ref{fig:R_Zmumu}. This plot indicates that there are many scenarios which can explain the muon $g-2$ anomaly within 1 sigma and simultaneously significantly  enhance $h \to 4 \mu$. Limiting Yukawa couplings to 0.5, the mass of $e_4$ has to be larger than about 113 GeV in order not to be ruled out by $h \to 4\mu$. Increasing the Yukawa coupling up to 1, the $m_{e_4}$can be close to the LEP limit for the small $M_L$ case. For the asymptotic case, $m_{e_4} $ is required to be larger than about 119 GeV. Selected contours of constant $R_{Z\mu\mu}$ from Fig.~\ref{fig:R_Zmumu} indicate the impact of improved limits in the future.

Predictions  for $h \to 2e2\mu$ and limits from the ATLAS analysis in this channel are almost identical and thus we do not show them separately. The comparison of the impact of ATLAS and CMS analyses is given in Appendix~\ref{app:CMS} for the $h\to 4\mu $ case. The CMS analysis does not separate $2e2\mu$ final states based on which pair of leptons originates from the on-shell $Z$, and thus the comparison in this channel is not possible.

In Appendix ~\ref{app:CMS} we also briefly discuss constraints from $h \to WW^*\to 2\ell 2\nu$ on $h \to e_4^\pm \mu^\mp \to W^\pm \mu^\mp \nu$. We show that these constraints are weaker than those from $h \to ZZ^*\to 4\ell$, unless BR($e_4^\pm \to W^\pm \nu$) is close to 1.

 \section{Discussion and Conclusions }
\label{sec:conclusions}

In extensions of the SM by vectorlike leptons the new lepton, $e_4$, may be lighter than the Higgs boson. Because of possible flavor violating couplings, $h \to e_4^\pm \mu^\mp$ with subsequent decays of the heavy lepton, $e_4^\pm \to Z\mu^\pm$, can significantly contribute to  $h \to Z\mu^+\mu^-$  and thus  affect the measurement of Higgs decays in $4\mu$ or $2e2\mu$ final states.

We derived limits on couplings and the mass of the new lepton from ATLAS and CMS searches for $h\to 4\ell$.  
We focused on the couplings of the muon and discussed possible effects on $h\to ZZ^* \to 4\mu$ or $2e2\mu$  from the region of parameters that can explain the anomaly in the measurement of the muon $g-2$.
Couplings required for the explanation of the muon $g-2$ also generate $h \to e_4^\pm \mu^\mp  \to Z\mu^+\mu^-$.  This scenario predicts equal enhancement of $h \to 4\mu$ and $2e2\mu$ (first pair of leptons originating from the on-shell Z)
and no enhancement in $h \to 4e$ and $2\mu2e$ final states. We showed that  scenarios that can explain the muon $g-2$ within $1\sigma$ are typically ruled out by $h \to 4\mu$ for  $m_{e_4} \lesssim 118$ GeV. However, there are some viable scenarios with  lighter $e_4$, even close to the LEP limit. 
We also showed the impact of improved limits on  $h\to ZZ^* \to 4\mu$ in future.  However, if $e_4$ is heavier than the Higgs boson, its effect on Higgs decay is negligible while it can still explain the muon $g-2$. We also derived constraints from $h \to WW^*\to 2\ell 2\nu$ on  $h \to e_4^\pm \mu^\mp \to W^\pm \mu^\mp \nu$. We showed that these constraints are weaker than those from $h \to ZZ^*\to 4\ell$, unless BR($e_4^\pm \to W^\pm \nu$) is close to 1.

Vectorlike leptons can be pair produced at the LHC. Although $e_4^\pm \to Z\mu^\pm$ provides a very clean and distinctive signal, the rate for this decay mode might be small. The remaining decay mode, $e_4^\pm \to W^\pm \nu$, which can be dominant, is harder to constrain. In addition, the new leptons can  also decay into $\tau$ leptons reducing the number of light leptons in final states. Compilation of constraints on vectorlike leptons from searches for anomalous production of multi lepton events can be found in Ref.~\cite{Dermisek:2014qca}.

\vspace{0.5cm}
\noindent
{\it Note added.} During the completion of this work papers studying effects of new physics on $h\to 4\ell$ appeared. In Ref.~\cite{Gonzalez-Alonso:2014rla},  authors studied effects of new light scalar or vector fields, 
and  in Ref.~\cite{Falkowski:2014ffa},   the  effects of light $SU(2)$ singlet vectorlike leptons were  also considered.

\acknowledgements


We would like to thank J. Hall and E. Lunghi for useful discussions. 
R.D. also thanks H.D. Kim for discussions and Seoul National University and  the Mainz Institute for Theoretical Physics (MITP) for their hospitality and  partial support during the completion of this work. S.S. thanks the Deutsches Elektronen-Synchrotron (DESY, Hamburg) for the support of visit during the process of this work. This work was supported in part by the Department of Energy under Grant NO. DE-FG02-13ER42002.

\appendix

\section{Limits on new physics contributing to $h  \to Z \mu^+\mu^-$ }
\label{app:limits}

In this Appendix we use ATLAS~\cite{atlas4l} and CMS~\cite{cms4l} searches for  $h \to ZZ^\ast \to 4\ell$ to derive the 95 \% C.L. upper limits on new physics contribution to $h  \to Z \mu^+\mu^-$  with $Z$ further decaying into $\mu^+ \mu^-$ or $e^+ e^-$. 
We adopt a modified frequentist construction (CL$_{\rm s}$)~\cite{cls} (see also Appendix D of \cite{Dermisek:2013cxa}).

The expected number of $h \to 4\ell$ events  after applying given selection cuts can be written as:
\begin{equation}
E = \eta \cdot \sigma^0 \cdot \mathcal{L} \\
= \frac{\eta}{\eta_{\rm SM}} \cdot \frac{\sigma^0}{\sigma^0_{\rm SM}} \cdot E_{\rm SM} \\
 = \xi \cdot \frac{\sigma^0}{\sigma^0_{\rm SM}} \cdot E_{\rm SM}~,
\end{equation}
where $\eta$ is the cut efficiency, $\sigma^0$ is the cross section that includes both the SM $h \to 4\ell$ and the new physics contribution before the cuts, and $\mathcal{L}$ is the integrated luminosity. The $\eta_{\rm SM}$, $\sigma_{\rm SM}^0$, and $E_{\rm SM}$ are the cut efficiency for the SM contribution only, the SM cross section before the cuts, and the expected number  of the SM $h \to ZZ^\ast \to 4\ell$ events, respectively. Finally,  $\xi$ is the relative cut acceptance $\eta / \eta_{\rm SM}$. The upper limit on $\sigma^0 / \sigma^0_{\rm SM}$ can be obtained by constraining the expected number of events $E$ to be  smaller than the 95 \% CL$_{\rm s}$ limit, $\ell_{95}$, obtained (below) from experiments,
\begin{equation}
\sigma^0 / \sigma^0_{\rm SM} <  \frac{\ell_{95}}{\xi \cdot E_{\rm SM}}~.
\label{eq:bound1}
\end{equation}
Equivalently, we can obtain a limit on 
\begin{equation}
\mu \equiv \sigma / \sigma_{\rm SM} = \xi \sigma^0/\sigma^0_{\rm SM}  < \frac{\ell_{95}}{E_{\rm SM}}~, 
\label{eq:bound2}
\end{equation}
where the cross sections $\sigma$ and  $\sigma_{\rm SM}$ are those after applying the cuts. This quantity is related to $R_{Z\mu\mu}$, defined in Eq.~(\ref{eq:R_Zmumu}):
\begin{equation}
R_{Z\mu\mu} 
= \xi \cdot \frac{\Gamma_h^{\rm tot}}{\Gamma_h^{\rm SM}} \cdot \frac{\sigma^0}{\sigma^0_{\rm SM}} = \mu  \cdot \frac{\Gamma_h^{\rm tot}}{\Gamma_h^{\rm SM}}
\end{equation}
in the region of the parameter space where the narrow width approximation  works well. Here, $\Gamma_h^{\rm tot}$ is the total decay width of the Higgs boson which includes possible decay mode to a new lepton, while $\Gamma_h^{\rm SM} = 4.07$ MeV is the SM expectation~\cite{Heinemeyer:2013tqa}.

In the modified frequentist construction, a confidence level $1-\alpha$ is obtained by the ratio of probabilities:
\begin{equation}
\alpha = \frac{P(D \ge \lambda|_{H_0})}{P(D \ge \lambda|_{H_1})}~,
\label{eq:cls}
\end{equation}
where $D$ is the data and $\lambda$ is the expected distribution in the signal-plus-background  ($H_0$) and in the background-only ($H_1$) hypotheses. The signal-plus-background is given by
\begin{equation}
S + B = s + b \pm \sqrt{s + \Sigma^2 s^2 + b+ \sigma_b^2}~,
\end{equation}
where $s$ is the expected  signal with  statistical uncertainty  $\sqrt{s}$. We assume that the fractional systematic uncertainties $\Sigma$ are the same as those in the SM $h \to ZZ^\ast \to 4\ell$ searches at 8 TeV.\footnote{For the 4$\mu$ final state these are 11.4\% at ATLAS~\cite{atlas4l} and 11.9\% at CMS~\cite{cms4l}. For $2e2\mu$ final state at ATLAS it is 11.6\%.} The  expected background is $b$ with  statistical uncertainty $\sqrt{b}$ and systematic uncertainty $\sigma_b$~\cite{atlas4l,cms4l}. Assuming every distribution is Gaussian we can take
\begin{eqnarray}
P(D \ge \lambda|_{H_0}) &= & \Phi\left( \frac{D-s-b}{\sqrt{s + \Sigma^2 s^2 +  b+ \sigma_b^2}}\right)~, \\
P(D \ge \lambda|_{H_1}) &=& \Phi\left( \frac{D-b}{\sqrt{b + \sigma_b^2}}\right)~, 
\end{eqnarray}
where $\Phi$ is the cumulative distribution function. For a given set of $D$, $b$, $\Sigma$, $\sigma_b$, and $\alpha$ the $1-\alpha$ confidence level limit is obtained from Eq.~(\ref{eq:cls}). Setting $\alpha = 0.05$ we obtain the 95 \% C.L. limit, $\ell_{95}$. 

In our analysis, in order to obtain the limits we include the SM $h \to ZZ^\ast \to 4\mu$ $(2e2\mu)$ process in the signal.
The upper bounds on $\mu = \sigma / \sigma_{\rm SM}$ as the 95\% C.L. limits from both  ATLAS and CMS data are shown in Table~I.
The ATLAS limits are from 4.6 and 20.7 fb$^{-1}$ at $\sqrt{s} = 7$ TeV and $\sqrt{s} = 8$ TeV respectively~\cite{atlas4l},  while the CMS limits are from 5.1 and 19.5 fb$^{-1}$ at $\sqrt{s} = 7$ TeV and $\sqrt{s} = 8$ TeV respectively~\cite{cms4l}. We do not include $2e2\mu$ final state for the CMS since they do not separate $2e2\mu$ and $2\mu2e$ data ($2e$ originating from on-shell or off-shell $Z$). 
\begin{table}[htdp]
\begin{center}
\caption{The 95\% CL$_{\rm s}$ exclusion bound on $\mu \equiv \sigma / \sigma_{\rm SM}$, which is $\ell_{95}/E_{\rm SM}$.}
\begin{tabular}{|c|c|c|}
\hline
  & ATLAS 7+8 TeV 4.6+20.7 fb$^{-1}$ & CMS 7+8 TeV 5.1 + 19.5 fb$^{-1}$ \\
\hhline{|=|=|=|}
$4\mu$ & 2.85& 1.85 \\
\hline
$2e2\mu$ & 3.02 & No separate $2e$ from on-shell $Z$ \\
\hline
\end{tabular}
\end{center}
\label{table:limit}
\end{table}

During the completion of this work a new ATLAS $h \to 4\ell$ analysis appeared~\cite{Aad:2014aba} based on 20.3 and 4.5 fb$^{-1}$ data at 8 and 7 TeV, respectively. Because of effective discrimination of $J/\psi \to \ell^+ \ell^-$ the number of observed events slightly increased. However, the bounds are similar to those in Table~I: 3.04 for $4\mu$ and 3.28 for $2e2\mu$.

\section{Details of the analysis and comparison of ATLAS and CMS}
\label{app:CMS}

\begin{figure}
\includegraphics[width=0.5\linewidth]{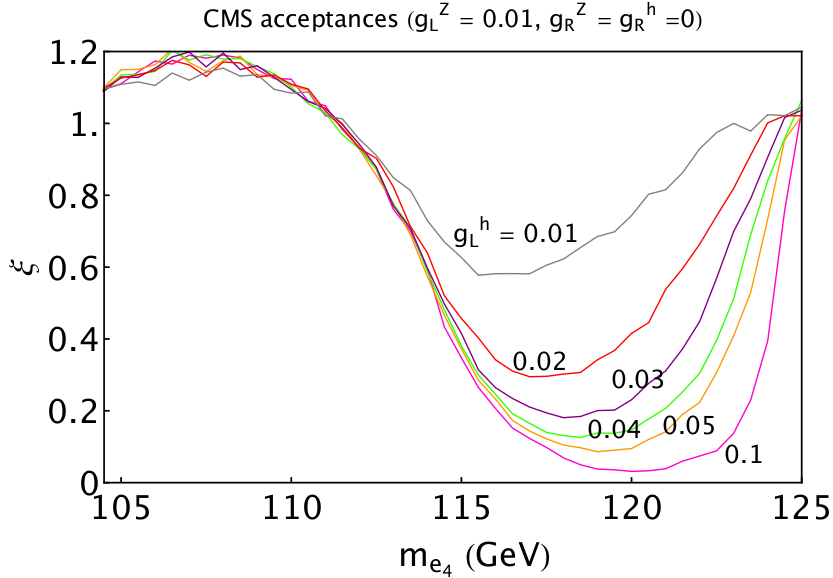}
\caption{The CMS acceptance of $h \to (e_4^\pm\mu^\mp,ZZ^*)\to 4\mu$ relative to the $h \to ZZ^*\to 4\mu$ in the SM.}
\label{fig:acc}
\end{figure}

The relative acceptances, $\xi$, are obtained using MadGraph 5~\cite{madgraph5} for simulating the process $g g \to h \to Z \mu^+ \mu^-$ with subsequent decay of $Z \to \mu^+ \mu^-$ or $e^+ e^-$ from the model written with FeynRules~\cite{feynrules}. We also used Pythia 6~\cite{Sjostrand:2006za} to include the initial and final state radiation. The production processes include the new physics $g g \to h \to e_4^\pm \mu^\mp \to 4\mu$ $ (2e2\mu)$, the SM $g g \to h \to ZZ^\ast \to 4\mu$ $ (2e2\mu)$, and their interference. The relative acceptances for the ATLAS analysis for selected values of couplings were given in Fig.~\ref{fig:acceptance} and, for comparison, the corresponding acceptances for the CMS analysis are given in Fig.~\ref{fig:acc}. In this figure we assume BR$(e_4^\pm \to Z \mu^\pm)$ = 1. The decay width of the Higgs boson is calculated including the new process $h \to e_4^\pm \mu^\mp$ .
The dependence of $\xi$ on the mass of the new lepton and the $g_L^h$ coupling is very similar to the ATLAS case.

\begin{figure}
\includegraphics[width=0.49\linewidth]{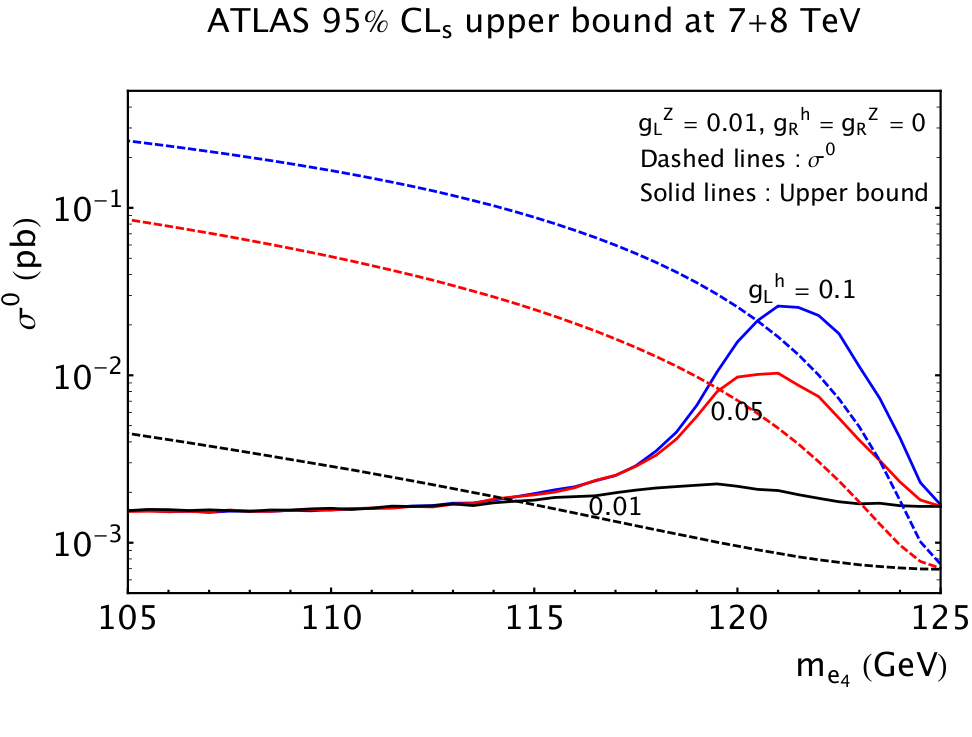}
\includegraphics[width=0.49\linewidth]{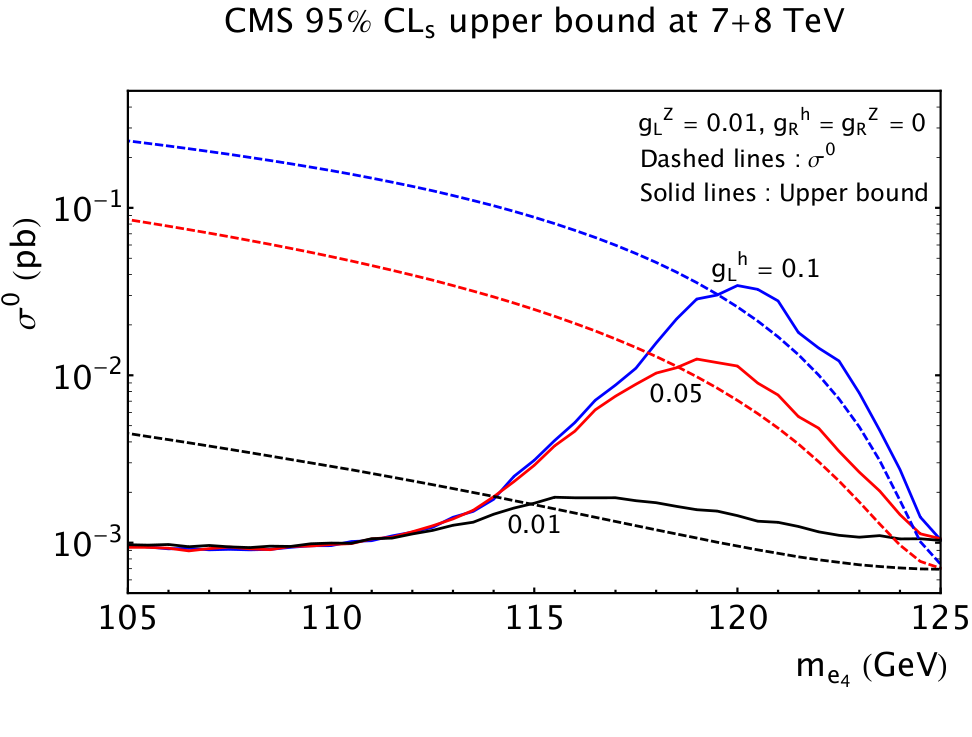}
\caption{Predicted cross section  $\sigma^0$ for  $gg \to h \to Z \mu^+ \mu^-  \to 4\mu$ as a function of $m_{e_4}$ for $g_L^h = 0.1, 0.05, 0.01$ (dashed lines from top to bottom) and with  $g_L^Z = 0.01, g_R^h = g_R^Z = 0$. The corresponding 95 \% C.L. limits from the ATLAS (left) and CMS (right) analyses are indicated by solid lines. }
\label{fig:boundsig0}
\end{figure}

The predicted cross section $\sigma^0$ for $4\mu$ final state is shown in Fig.~\ref{fig:boundsig0} as a function of $m_{e_4}$ for three choices of couplings $g_L^h = 0.1, 0.05, 0.01$ with  $g_L^Z = 0.01, g_R^h = g_R^Z = 0$. It is assumed that BR$(e_4^\pm \to Z \mu^\pm)$ = 1.  
We used the $k$ factor 2.62 in order to match the leading order result for $\sigma(g g \to h)$ obtained from MadGraph 5 with  the value presented by the Higgs Working Group~\cite{Heinemeyer:2013tqa} which includes higher order contributions. We use the same factor to  multiply the cross section for $g g \to h \to Z \mu^+ \mu^- \to 4\mu$ obtained from  MadGraph 5 now  including the contribution from the new lepton. The cross section of the SM $h \to ZZ^\ast \to 4\mu$ process $\sigma_{\rm SM}^0$ before applying the cuts is obtained from Ref.~\cite{Heinemeyer:2013tqa}. The derived 95 \% C.L. upper limits on $\sigma^0$ from the ATLAS (left) and CMS (right) analyses for each choice of the $g_L^h$ coupling are indicated by solid lines in Fig.~\ref{fig:boundsig0}. The bounds are inversely proportional to $\xi$. 

\begin{figure}
\includegraphics[width=0.49\linewidth]{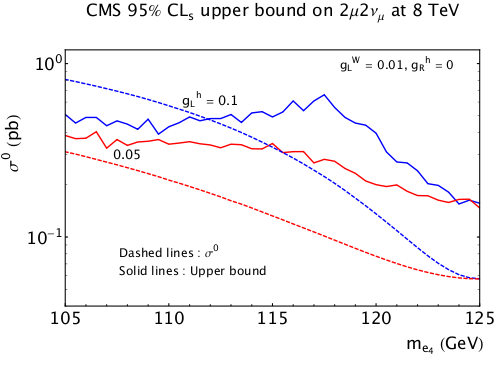}
\includegraphics[width=0.49\linewidth]{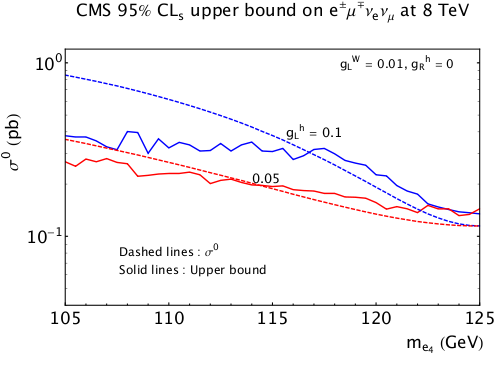}
\caption{Predicted cross sections $\sigma^0$ for $gg \to h \to W^\pm \mu^\mp \nu_\mu \to \mu^+ \mu^- 2\nu_\mu$ (left) and $e^\pm \mu^\mp \nu_e \nu_\mu$ (right) as functions of $m_{e_4}$ for $g_L^h = 0.1, 0.05$ (dashed lines from top to bottom) and with  $g_L^W = 0.01, g_R^h  = 0$. The corresponding 95 \% C.L. limits from the  CMS analysis are indicated by solid lines as discussed in detail in Ref.~\cite{Dermisek:2013cxa}.}
\label{fig:boundsig0ww}
\end{figure}

Finally, we also analyzed the impact of constraints from $h \to WW^*\to 2\ell 2\nu$  on $h \to e_4^\pm \mu^\mp \to W^\pm \mu^\mp \nu$.
The predicted cross sections for $gg \to h \to W^\pm \mu^\mp \nu_\mu \to \mu^+ \mu^- 2\nu_\mu$ and $e^\pm \mu^\mp \nu_e \nu_\mu$ as functions of $m_{e_4}$ for selected values of $g_L^h$ are given in Fig.~\ref{fig:boundsig0ww}. The corresponding 95 \% C.L. limits from the  CMS analysis \cite{CMS_WW}  are indicated by solid lines. They are obtained following the procedure discussed in detail in Ref.~\cite{Dermisek:2013cxa}. 
For each final state the events are further categorized according to the number of external jets mainly from the initial state radiations. Among them the strongest bound is selected for each value of $m_{e_4}$.

\begin{figure}
\includegraphics[width=0.5\linewidth]{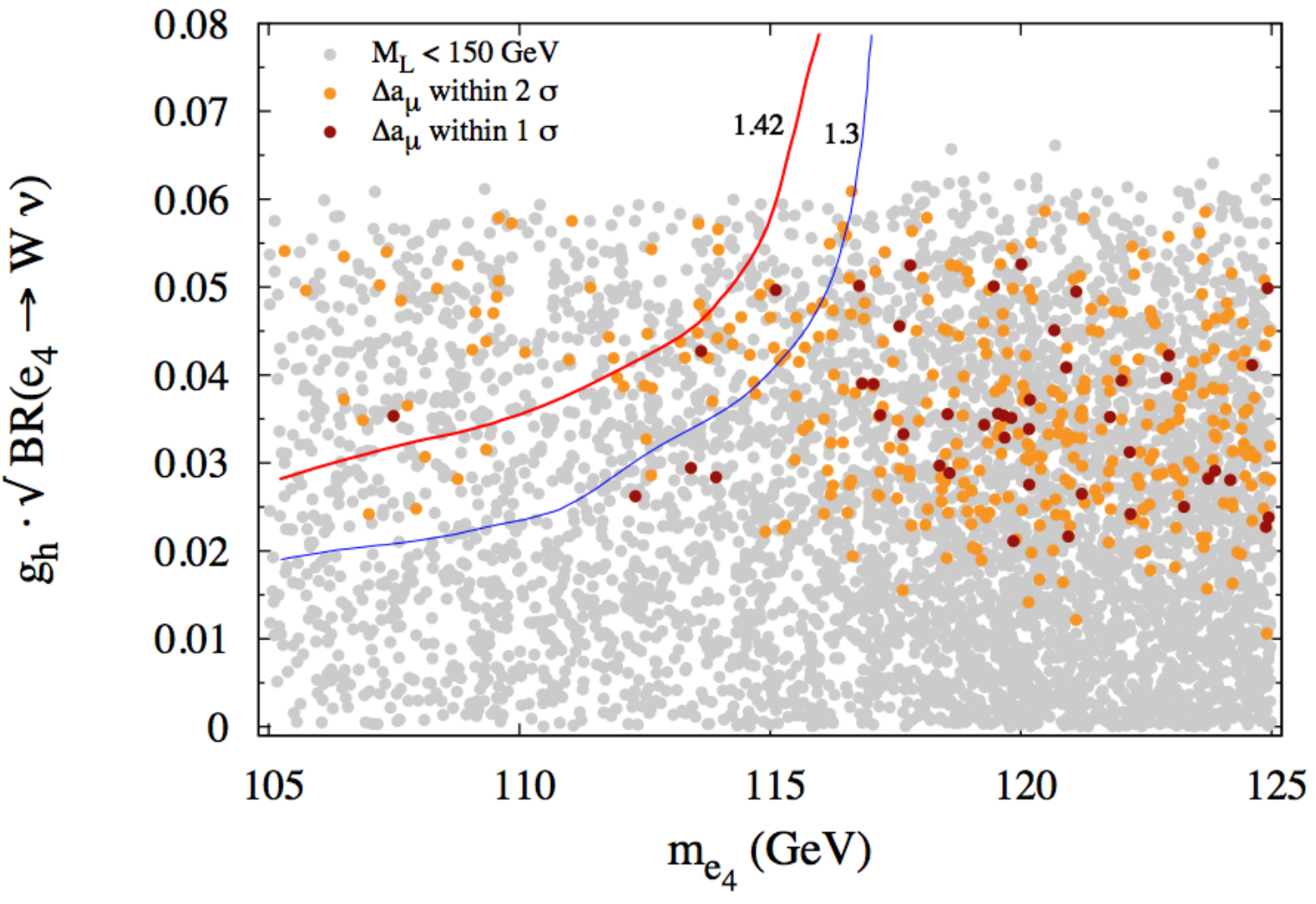}
\caption{Subset of points from Fig.~\ref{fig:1p1p_gh_me4} (left) that satisfy the ATLAS limit  $R_{Z\mu\mu} < 2.8$  plotted in $g_h \sqrt{{\rm BR}(e_4\to W\nu)}$ -- $m_{e_4}$ plane.  The thick (red) line indicates  the
limit on  $\mu \equiv \sigma / \sigma_{\rm SM}$ for $e^\pm \mu^\mp \nu_e \nu_\mu$ final state. In order to illustrate the impact of improved limits in the future we also show the contour of $\mu = 1.3$. 
}
\label{fig:hwwcontour}
\end{figure}

The constraints from $h \to WW^*\to 2\ell 2\nu$ on the new vectorlike lepton  are typically weaker than those from $h \to ZZ^*\to 4\ell$ , unless BR($e_4^\pm \to W^\pm \nu$) is close to 1. Thus these constraints are relevant only in a limited region of the parameter space. This is illustrated in Fig.~\ref{fig:hwwcontour} where we plot the points from Fig.~\ref{fig:1p1p_gh_me4} (left) that satisfy the ATLAS limit  $R_{Z\mu\mu} < 2.8$ in $g_h \sqrt{{\rm BR}(e_4\to W\nu)}$ -- $m_{e_4}$ plane. The approximate limit on  $\mu \equiv \sigma / \sigma_{\rm SM}$ for $e^\pm \mu^\mp \nu_e \nu_\mu$ final state (which is stronger than the limit for $2\mu 2\nu_\mu$ final state)  is indicated by a thick (red) line. Only few points not excluded by $h \to ZZ^*\to 4\ell$ are excluded by  $h \to WW^*\to 2\ell 2\nu$.

\section{Partial width of  $h \to Z f_1 f_2$}
\label{app:width}

\small

\begin{center}
\begin{figure}[h]
 \includegraphics[width=12cm]{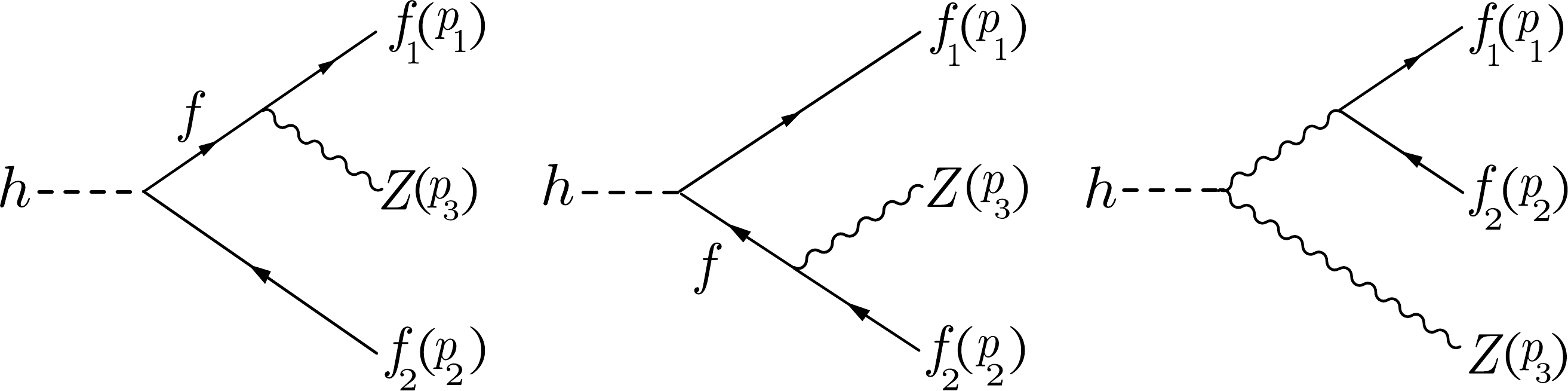}
 \caption{Tree level Feynman diagrams for $h \rightarrow Z f_1 f_2$.}
 \label{fig:digdigdig}
 \end{figure}
\end{center}

In this Appendix we calculate the partial decay width of a scalar $h$ decaying into $Z f_1 f_2$  mediated by a new fermion $f$ and the $Z^*$.  This calculation can be straightforwardly modified for  $h \to W f_1 f_2$. The relevant Feynman diagrams are shown in Fig.~\ref{fig:digdigdig}. 

\subsection{Amplitudes}

The amplitudes for the three diagrams are: 
\small
\begin{eqnarray}
iT_1 &=&  \bar{u}_{f_1} ({\bf p_1})\frac{ig}{2c_W} \gamma^\mu (c_{v1}  + c_{a1}  \gamma_5)  \frac{1}{i} \frac{\slashed{p}_1 + \slashed{p}_3 - m_f}{(p_1 + p_3)^2 + m_f^2 - i\Gamma_f m_f} \frac{-i(y_{v1} + y_{a1}\gamma_5)}{2\sqrt{2}} v_{f_2}({\bf p_2}) \epsilon_\mu, \nonumber \\
\nonumber \\
iT_2 &=&  \bar{u}_{f_1} ({\bf p_1}) \frac{-i(y_{v2} + y_{a2}\gamma_5)}{2\sqrt{2}}  \frac{1}{i} \frac{\slashed{p}_2 + \slashed{p}_3 - m_f}{(p_2 + p_3)^2 + m_f^2 - i\Gamma_f m_f} \gamma^\mu  \frac{ig}{2c_W}(c_{v2}  + c_{a2}  \gamma_5)  v_{f_2}({\bf p_2}) \epsilon_\mu, \nonumber \\
\nonumber \\
iT_Z &=&  \frac{-igM_Z}{c_W}\bar{u}_{f_1}({\bf p_1})\gamma^\mu\frac{ig}{2c_W}(c_v + c_a\gamma_5)v_{f_2}({\bf p_2})
\frac{1}{i}\frac{g_{\mu\nu}}{(p_1 + p_2)^2 + M_Z^2 - i \Gamma_Z M_Z}\epsilon^{\nu}(p_3),
\end{eqnarray}
where the notation for the couplings of the $Z$ boson to fermions $f$ and $f_i$ follows from the Lagrangian, $\mathcal{L} \supset (g/2c_W) \bar{f}_i  \gamma^\mu (c_{vi}  + c_{ai}  \gamma_5)f Z_\mu  + h.c.$, and the couplings of the scalar are defined by $\mathcal{L} \supset   -  1/(2\sqrt{2})  \bar{f}_i\gamma^\mu (y_{vi}  + y_{ai}  \gamma_5) f h + h.c.$.

Writing $k_{1,2} \equiv p_{1,2} + p_3$, the amplitude squared for the first diagram is given by
\small
\begin{eqnarray}
|T_1|^2 &=&\frac{1}{8} \frac{g^2}{4c_W^2 } \frac{1}{(k_1^2 + m_f^2)^2 + \Gamma_f^2 m_f^2}\epsilon_\mu (p_3) \epsilon_\nu^* (p_3)\nonumber\\
&&\text{Tr}\left[u_{f_1} \bar{u}_{f_1} \gamma^\mu (c_{v1} +c_{a1} \gamma_5)  (\slashed{k}_1 - m_f) (y_{v1} + y_{a1}\gamma_5) v_{f_2} \bar{v}_{f_2} \nonumber
                        (y_{v1} - y_{a1}\gamma_5)(\slashed{k}_1 - m_f) \gamma^\nu  (c_{v1} +c_{a1} \gamma_5) \right]   .
\end{eqnarray}
The polarization vectors $\epsilon_\mu$ of the $Z$ boson satisfy
\small
\begin{equation}
 \sum_{\lambda = \pm,0} \epsilon_\lambda^{*\rho}(k) \epsilon_\lambda^\sigma(k) = g^{\rho\sigma} + \frac{k^\rho k^\sigma}{M_Z^2},
\end{equation}
and summing over final spins, we get
\begin{eqnarray}
\langle |T_1|^2 \rangle &=& \sum_{\lambda = \pm,0} \sum_{f_1, f_2 = \pm} |T_1|^2 =  \frac{1}{8} \frac{g^2}{4c_W^2} \frac{1}{(k_1^2 + m_f^2)^2+ \Gamma_f^2 m_f^2}\left( g_{\mu\nu} + \frac{{p_3}_\mu {p_3}_\nu}{M_Z^2}     \right)   \mathcal{M}_1^{\mu\nu} \label{t1sq}, \\ \nonumber
\end{eqnarray}
where
 \begin{eqnarray}
 \mathcal{M}_1^{\mu\nu} &=&\text{Tr}[(-\slashed{p}_1 + m_1)\gamma^\mu (c_{v1} +c_{a1} \gamma_5)  (\slashed{k}_1 - m_f) (y_{v1} + y_{a1}\gamma_5) \nonumber \\
&&\phantom{aaaaaaaaaaaaaaaaa}    (-\slashed{p}_2 - m_2) (y_{v1} - y_{a1}\gamma_5)(\slashed{k}_1 - m_f) \gamma^\nu  (c_{v1} +c_{a1} \gamma_5) ] .
\end{eqnarray}

The $\langle |T_2|^2 \rangle$ can be obtained from $\langle |T_1|^2 \rangle$, by making the following replacements: 
\begin{itemize}
\item $p_1 \leftrightarrow p_2$ and $m_1 \leftrightarrow -m_2$,
\item $c_{v1} \rightarrow c_{v2}$, $c_{a1} \rightarrow c_{a2}$, $y_{v1} \rightarrow y_{v2}$ and $y_{a1} \rightarrow -y_{a2}$.
\end{itemize}
\vspace{.25cm}
%
%
Interference between the two diagrams is given by twice the real part of 	
\small
\begin{eqnarray}
\langle T_1\bar{T}_2 \rangle &=& \sum_{\lambda = \pm,0} \sum_{f_1, f_2 = \pm} T_1\bar{T}_2 \nonumber \\
&=&  \frac{1}{8}\frac{g^2}{4c_W^2}\left(\frac{1}{k_1^2 + m_f^2- i\Gamma_f m_f}\right)  \left(\frac{1}{k_2^2 + m_f^2- i\Gamma_f m_f}\right)^*  \left( g_{\mu\nu} + \frac{{p_3}_\mu {p_3}_\nu}{M_Z^2}     \right) \mathcal{M}_{12}^{\mu\nu} ,  \nonumber 
\end{eqnarray}
 where
  \begin{eqnarray}
 \mathcal{M}_{12}^{\mu\nu} &=& \text{Tr}[(-\slashed{p}_1 + m_1) \gamma^\mu (c_{v1} +c_{a1} \gamma_5) (\slashed{k}_1 - m_f) (y_{v1} + y_{a1}\gamma_5)  \nonumber \\
 &&\phantom{aaaaaaaaaaaaaaaa}          (-\slashed{p}_2 - m_2) \gamma^\nu (c_{v2}  + c_{a2}  \gamma_5)  (\slashed{k}_2 - m_f) (y_{v2} - y_{a2}\gamma_5) ] .
\end{eqnarray}

%
%
The interference with $ZZ^*$ is given by 2Re$\langle T_1\bar{T}_Z \rangle$ + 2Re$\langle T_2\bar{T}_Z \rangle$, where
\small
\begin{eqnarray}  
\langle T_1\bar{T}_Z \rangle &=& - \frac{g^3 M_Z}{8 \sqrt{2}c_W^3} \frac{1}{k_1^2 + m_f^2 -i \Gamma_f m_f} \left( \frac{1}{(p_1 + p_2)^2 + M_Z^2 -i \Gamma_Z M_Z}\right)^* \left(g_{\mu\nu} + \frac{{p_3}_\mu{p_3}_\nu}{M_Z^2}\right)  \mathcal{M}_{1Z}^{\mu\nu}\nonumber 
\end{eqnarray}
with
\begin{eqnarray} 
 \mathcal{M}_{1Z}^{\mu\nu}&=& \text{Tr}[(-\slashed{p}_1 + m_1) \gamma^\mu (c_{v1} + c_{a1}\gamma_5) (\slashed{k}_1 - m_f)(y_{v1} + y_{a1}\gamma_5)(-\slashed{p}_2 - m_2) \gamma^\nu (c_{v} + c_{a}\gamma_5)] ,
\end{eqnarray}
%
and
$\langle T_2\bar{T}_Z \rangle$ can be obtained from $\langle T_1\bar{T}_Z \rangle$ by making the same substitutions as for  $\langle |T_2|^2 \rangle$, listed above.
%

\subsection{Traces} 
The results for the traces appearing in the  amplitudes squared are:
\small
\begin{eqnarray}
\mathcal{M}^{\mu\nu}_1 g_{\mu\nu} &=& 8 ({c_{v1} }^2 + {c_{a1} }^2) (y_{v1}^2 + y_{a1}^2) \left( 2 (p_1 \phantom{}k_1)(k_1\phantom{}p_2) - (p_1 \phantom{}p_2)k_1^2 - m_f^2  ( p_1 \phantom{}p_2)  \right)    \nonumber \\   
&+&  16 m_f m_2 ({c_{v1} }^2 + {c_{a1} }^2) (y_{v1}^2 - y_{a1}^2) \left( p_1 \phantom{}k_1 \right) + 32 m_1 m_f ({c_{v1} }^2 - {c_{a1} }^2) (y_{v1}^2+y_{a1}^2) ( k_1 \phantom{}p_2)   \nonumber \\   
&-&  16 m_1 m_2 ({c_{v1} }^2 - {c_{a1} }^2)( y_{v1}^2 - y_{a1}^2) (k_1^2 - m_f^2 ) \nonumber \\
& -&  32 c_{v1} c_{a1}  y_{v1}y_{a1} \left( 2 (p_1 \phantom{}k_1)(k_1\phantom{}p_2) - (p_1 \phantom{}p_2)k_1^2 + m_f^2 (p_1 \phantom{}p_2) \right) , \label{Pdig1} \\
\nonumber\\
\mathcal{M}^{\mu\nu}_1 {{p_3}_\mu {p_3}_\nu} &=& 4(({c_{v1} }^2 + {c_{a1} }^2)(y_{v1}^2 + y_{a1}^2) - 4 c_{v1} c_{a1}  y_{v1}y_{a1}) \nonumber \\
&&\phantom{aaaaaa}\{ -4(p_1 \phantom{}p_3)(k_1 \phantom{}p_2)(k_1 \phantom{}p_3) + 2(p_1 \phantom{}p_3) k_1^2 (p_2 \phantom{}p_3) -2(p_1 \phantom{}k_1) (p_2\phantom{}k_1)M_Z^2  +(p_1\phantom{}p_2) k_1^2 M_Z^2\} \nonumber \\
&+& 4 m_f^2 (({c_{v1} }^2 + {c_{a1} }^2)(y_{v1}^2 + y_{a1}^2) + 4 c_{v1} c_{a1}  y_{v1}y_{a1}) \left\{2(p_1 \phantom{}p_3)(p_3 \phantom{}p_2) + (p_1 \phantom{}p_2)M_Z^2 \right\} \nonumber \\
&-&  8 m_f  m_2 ({c_{v1} }^2 + {c_{a1} }^2)  (y_{v1}^2 - y_{a1}^2) \left\{ 2(p_1 \phantom{}p_3)(p_3 \phantom{}k_1) +(p_1 \phantom{}k_1)M_Z^2 \right\} \nonumber \\
&+&  4 m_1 m_2 M_Z^2({c_{v1} }^2 - {c_{a1} }^2) (y_{v1}^2 - y_{a1}^2)   k_1^2  - 8 m_1 m_f M_Z^2 ({c_{v1} }^2 - {c_{a1} }^2) (y_{v1}^2 + y_{a1}^2)( p_2 \phantom{}k_1) \nonumber \\ 
&-&     4 m_1 m_2 m_f^2M_Z^2  ({c_{v1} }^2 - {c_{a1} }^2) (y_{v1}^2 - y_{a1}^2) , \label{trace1}
\end{eqnarray}
\begin{eqnarray}
\mathcal{M}^{\mu\nu}_{12} g_{\mu\nu} &=& 16 (c_{v1}c_{v2} - c_{a1}c_{a2}) y_{v1}y_{v2}((k_1 \phantom{}p_2) + m_f m_2)  \left(-p_1 \phantom{}k_2 + m_1 m_f\right) \nonumber\\
&& \phantom{aa}- 8 (c_{v1} c_{v2} + c_{a1} c_{a2}) y_{v1}y_{v2} \left(-m_1m_2  (k_1 \phantom{}k_2) + m_1 m_f (p_2 \phantom{}k_2)
                               - m_2m_f (k_1 \phantom{}p_1) + m_f^2(p_1 \phantom{}p_2) \right) \nonumber\\
&& -16 (c_{v1}c_{v2} - c_{a1}c_{a2}) y_{a1}y_{a2}((k_1 \phantom{}p_2) -  m_f m_2) \left((p_1 \phantom{}k_2) + m_1m_f\right) \nonumber\\
&& \phantom{aa}+ 8 (c_{v1} c_{v2} + c_{a1} c_{a2})  y_{a1}y_{a2} \left(-m_2 m_1(k_1 \phantom{}k_2) - m_f m_1(p_2 \phantom{}k_2)
                              +m_f m_2 (p_1 \phantom{}k_1) + m_f^2(p_1 \phantom{}p_2) \right), \nonumber\\ 
&& - 16  (c_{v1}c_{a2} - c_{a1}c_{v2})(y_{v1}y_{a2} + y_{a1} y_{v2})  \left( (p_1 \phantom{}k_2) (k_1 \phantom{}p_2) + m_1 m_2 m_f^2 \right) \nonumber\\
&& -8 (c_{v1}c_{a2} + c_{a1}c_{v2})(y_{v1}y_{a2} + y_{a1} y_{v2})  m_f  \left(m_1(p_2 \phantom{}k_2) + m_2 (p_1 \phantom{}k_1) \right)            \nonumber \\
&& -8(c_{v1}c_{a2} + c_{a1}c_{v2})(y_{v1}y_{a2} - y_{a1} y_{v2})\left(m_1m_2(k_1 \phantom{}k_2) - m_f^2 (p_1 \phantom{}p_2) \right)      \nonumber \\     
&& +8(c_{v1}c_{a2} - c_{a1}c_{v2})(y_{v1}y_{a2} - y_{a1} y_{v2})\left(m_fm_2(p_1 \phantom{}k_2) + m_1m_f (k_1 \phantom{}p_2) \right)          ,                
\end{eqnarray}
\begin{eqnarray}
\mathcal{M}^{\mu\nu}_{12} {{p_3}_\mu {p_3}_\nu} 
&=&  4 \left((c_{v1} c_{v2} - c_{a1} c_{a2})(y_{v1}y_{v2} +  y_{a1}y_{a2}) + (c_{v1}c_{a2} - c_{a1}c_{v2} )(y_{v1}y_{a2} + y_{a1} y_{v2}) \right) \nonumber\\
&&\phantom{aa} \{ ({k}_2\phantom{}{p}_1)(k_1 \phantom{}p_2)M_Z^2 - 2 ({k}_2\phantom{}{p}_3)( 2(p_1 \phantom{}p_2)(k_1 \phantom{}p_3) - (p_1 \phantom{}k_1)(p_2\phantom{}p_3))\nonumber \\
&&\phantom{aa}- ({k}_2\phantom{}{k}_1)(2(p_1 \phantom{}p_3)(p_3 \phantom{}p_2) + (p_1 \phantom{}p_2)M_Z^2 ) + ({k}_2 \phantom{}{p}_2)(2(p_1 \phantom{}p_3)(p_3 \phantom{}k_1) + (p_1 \phantom{}k_1)M_Z^2) \} \nonumber\\ 
&-&  4  \left( (c_{v1}c_{v2} + c_{a1}c_{a2})(y_{v1}y_{v2} +  y_{a1}y_{a2})  -  (c_{v1}c_{a2} + c_{a1}c_{v2} ) (y_{v1}y_{a2} + y_{a1} y_{v2}) \right)  \nonumber\\
&&\phantom{aaaaaaaaaaaaaaaaaaaaaaaaaaa}m_f m_2\left(2(p_1 \phantom{}p_3)(p_3 \phantom{}k_1) +  (p_1 \phantom{}k_1)M_Z^2 \right)\nonumber\\
&+&  4 \left((c_{v1}c_{v2} + c_{a1}c_{a2})(y_{v1}y_{v2} -  y_{a1}y_{a2}) - (c_{v1}c_{a2} + c_{a1}c_{v2})(y_{v1}y_{a2} - y_{a1}y_{v2}) \right) \nonumber\\
&&\phantom{aaaaaaaaaaaaaaaaaaaaaaaaaaa} m_f^2\left(2(p_1 \phantom{}p_3)(p_3 \phantom{}p_2) + (p_1 \phantom{}p_2)M_Z^2 \right)  \nonumber\\
&+&  4 \left((c_{v1}c_{v2} - c_{a1}c_{a2})(y_{v1}y_{v2} -  y_{a1}y_{a2}) - (c_{v1}c_{a2} - c_{a1}c_{v2})(y_{v1}y_{a2} - y_{a1}y_{v2})  \right)  \nonumber\\
&&\phantom{aaaaaaaaaaaaaaaaaaaaaaaaaaa}M_Z^2m_f m_2 (p_1 \phantom{}k_2)\nonumber\\
&-&   4 \left((c_{v1}c_{v2} - c_{a1}c_{a2})(y_{v1}y_{v2} -  y_{a1}y_{a2}) - (c_{v1}c_{a2} - c_{a1}c_{v2})(y_{v1}y_{a2} - y_{a1}y_{v2})  \right)   \nonumber\\
&&\phantom{aaaaaaaaaaaaaaaaaaaaaaaaaaa}M_Z^2m_f m_1(k_1 \phantom{}p_2)\nonumber \\
&-&   4  \left((c_{v1}c_{v2} + c_{a1}c_{a2})(y_{v1}y_{v2} -  y_{a1}y_{a2})  - (c_{v1}c_{a2} + c_{a1}c_{v2})(y_{v1}y_{a2} - y_{a1}y_{v2})\right)  \nonumber \\
&&\phantom{aaaaaaaaaaaaaaaaaaaaaaaaaaa}m_1m_2\left(2(p_3 \phantom{}k_2)(p_3 \phantom{}k_1) + M_Z^2(k_1 \phantom{}k_2) \right) \nonumber \\
&+& 4   \left( (c_{v1}c_{v2} + c_{a1}c_{a2})(y_{v1}y_{v2} +  y_{a1}y_{a2})  + (c_{v1}c_{a2} + c_{a1}c_{v2} ) (y_{v1}y_{a2} + y_{a1} y_{v2}) \right) \nonumber \\
&&\phantom{aaaaaaaaaaaaaaaaaaaaaaaaaaa} m_1m_f \left(2(p_3 \phantom{}k_2)(p_3 \phantom{}p_2)+ M_Z^2(p_2 \phantom{}k_2) \right) \nonumber \\
&-&  4  \left( (c_{v1}c_{v2} - c_{a1}c_{a2})(y_{v1}y_{v2} +  y_{a1}y_{a2}) -(c_{v1}c_{a2} - c_{a1}c_{v2} )(y_{v1}y_{a2} + y_{a1} y_{v2}) \right) \nonumber\\ 
&&\phantom{aaaaaaaaaaaaaaaaaaaaaaaaaaa}m_1 m_2 m_f^2 M_Z^2,
\end{eqnarray}
%
%
\begin{eqnarray}
\mathcal{M}^{\mu\nu}_{1Z}g_{\mu\nu}&=& -8c_{v} y_{v1}  c_{v1} (2 m_1 (k_1 \phantom{}p_2) + 2 m_1 m_2 m_f + m_2 (p_1 \phantom{}k_1) - m_f (p_1 \phantom{}p_2)) \nonumber\\
 & + &   8c_{a} y_{v1} c_{a1}  (2 m_1 (k_1 \phantom{}p_2)+ 2 m_1 m_2 m_f - m_2 (p_1 \phantom{}k_1) + m_f (p_1 \phantom{}p_2)) \nonumber\\
 & - &   8c_{a} y_{a1} c_{v1}   (2 m_1 (k_1 \phantom{}p_2) - 2 m_1 m_2 m_f - m_2 (p_1 \phantom{}k_1) - m_f (p_1 \phantom{}p_2))\nonumber \\ 
 & + &   8c_{v} y_{a1} c_{a1}  (2 m_1 (k_1 \phantom{}p_2)- 2 m_1 m_2 m_f + m_2 (p_1 \phantom{}k_1) + m_f (p_1 \phantom{}p_2)) ,
\end{eqnarray}

\begin{eqnarray}
&&\mathcal{M}^{\mu\nu}_{1Z} {{p_3}_\mu {p_3}_\nu}  = \nonumber \\
&& 4c_{v}c_{v1}y_{v1}  \left (m_1 M_Z^2(k_1 \phantom{}p_2 +m_2m_f) + m_2(2 (p_3\phantom{}k_1) (p_1\phantom{}p_3) + (p_1 \phantom{}k_1)M_Z^2) - m_f(2 (p_1\phantom{}p_3) (p_2\phantom{}p_3) + p_1 \phantom{}p_2M_Z^2)\right) \nonumber\\ 
 &-&  4 c_{a}c_{a1} y_{v1}\left(m_1 M_Z^2(k_1 \phantom{}p_2+ m_2m_f) - m_2(2 (p_3\phantom{}k_1) (p_1\phantom{}p_3) + (p_1 \phantom{}k_1)M_Z^2) + m_f(2 (p_1\phantom{}p_3) (p_2\phantom{}p_3) + p_1 \phantom{}p_2M_Z^2)\right) \nonumber\\
 &+& 4 c_{a}c_{v1}y_{a1}\left(m_1M_Z^2(k_1 \phantom{}p_2 - m_2m_f) - m_2(2 (p_3\phantom{}k_1) (p_1\phantom{}p_3) + (p_1 \phantom{}k_1)M_Z^2) - m_f(2 (p_1\phantom{}p_3) (p_2\phantom{}p_3) + p_1 \phantom{}p_2M_Z^2)\right) \nonumber\\ 
 &-& 4c_{v}c_{a1} y_{a1} \left (m_1M_Z^2 (k_1 \phantom{}p_2 - m_2m_f) + m_2(2 (p_3\phantom{}k_1) (p_1\phantom{}p_3) + (p_1 \phantom{}k_1)M_Z^2) + m_f(2 (p_1\phantom{}p_3) (p_2\phantom{}p_3) + p_1 \phantom{}p_2M_Z^2)\right). \nonumber \\
\end{eqnarray}

%
%

\subsection{Kinematics}

The higgs 4-momentum is denoted by $P = p_1 + p_2 + p_3$. Defining $s_i = -(P-p_i)^2$ for i=1 to 3, we also have $\sum_i s_i = M_h^2 + M_Z^2 + m_1^2 + m_2^2$. Therefore the scalar products of the momenta are: 
\begin{eqnarray}
2p_1p_2&=& s_1 + s_2 - M_h^2 - M_Z^2, \nonumber\\ 
2p_1p_3&=& -s_2 + m_1^2 + M_Z^2, \nonumber\\ 
2p_2p_3&=& -s_1 + m_2^2 + M_Z^2 .
\end{eqnarray}
Substituting these into the amplitudes squared, we can integrate over $s_{1,2}$. The decay rate is given by
\begin{eqnarray}
\Gamma &=& \frac{1}{(2\pi)^3} \frac{1}{32 M_h^3} \int_{s_1^-}^{s_1^+} \int_{s_2^-}^{s_2^+} \langle|T|^2\rangle \; ds_1 ds_2 , \label{mygamma}
\end{eqnarray}
with the limits defined as in \cite{pdg}, \cite{rel_kin}:
\small
\begin{eqnarray}
s_1^-  &=& (m_2 + M_Z)^2 ,\nonumber \\
s_1^+ &=& (M_h - m_1)^2 ,\nonumber\\
s_2^-  &=& m_1^2 + M_Z^2 + \frac{1}{2s_1}((M_h^2 - s_1 - m_1^2)(s_1 - m_2^2 + M_Z^2) - \lambda(s_1, M_h^2, m_1^2)\lambda(s_1, m_2^2, M_Z^2)),\nonumber\\ 
s_2^+ &=& m_1^2 + M_Z^2 +\frac{1}{2s_1}((M_h^2 - s_1 - m_1^2)(s_1 - m_2^2 + M_Z^2) + \lambda(s_1, M_h^2, m_1^2)\lambda(s_1, m_2^2, M_Z^2)),\nonumber 
\end{eqnarray}
where $\lambda(x,y,z) = (x^2 + y^2 + z^2 - 2xy - 2yz - 2zx)^{1/2} $.

The $h\to Z\mu^+\mu^-$ can be obtained from the formulas above  
 by substituting $f_1 = f_2 = \mu$,  $f = e_{4}$, and 
\begin{eqnarray*}
c_{v1} = c_{v2} &=& \frac{c_W}{g}(g_L^Z + g_R^Z),   \\ 
c_{a1} = c_{a2} &=& \frac{c_W}{g}(-g_L^Z + g_R^Z),  \\ 
y_{v1} = y_{v2} &=&   g_L^h + g_R^h,  \\ 
y_{a1} = y_{a2} &=& -g_L^h + g_R^h  ,
\end{eqnarray*}
to match the notation in Eq.~(\ref{eq:eff_lagrangian}).

\subsection{$h \rightarrow Wtb$}
As a special case, in order to check and illustrate the usefulness of the general formulas, we calculate the partial decay width of the Higgs boson into $Wtb$ in the SM. There is only one diagram in this case since the contribution from $WW^*$ is negligible. 
To get this result from our calculation, we make the following replacements:
\begin{itemize}
\item $M_Z \rightarrow M_W$,
\item $m_{1,2} \rightarrow m_{b,t}$ and $m_f = m_t$,
\item $\frac{g}{2c_W} \rightarrow \frac{g}{\sqrt{2}}$ and $c_{v1} = -c_{a1} = 1/2$,
\item $\frac{y_{v1}}{2\sqrt{2}} \rightarrow  \frac{m_t}{v}$ and  $\frac{y_{a1}}{2\sqrt{2}} \rightarrow 0$,
\end{itemize}
where v = 174 GeV. The amplitude squared is 
\small
\begin{eqnarray}
\langle |T_1|^2 \rangle &=& \frac{g^2 m_t^2}{2 v^2}  \frac{1}{(-s_2 + m_t^2)^2 + \Gamma_t^2 m_t^2}\left( g_{\mu\nu} + \frac{{p_3}_\mu {p_3}_\nu}{M_W^2}     \right)   \mathcal{M}_1^{\mu\nu}
\equiv \frac{g^2m_t^2 }{2v^2} \frac{ M_h^2}{ M_W^2} \frac{\Gamma_0}{ y_t^2 + \gamma_t \kappa_t},
\end{eqnarray}
where we have defined: $\Gamma_0 \equiv \left( \kappa_W \frac{ g_{\mu\nu}}{M_h^4} + \frac{{p_3}_\mu {p_3}_\nu}{M_h^6}\right) \mathcal{M}_1^{\mu\nu} $,
 $x_{t,b} = \frac{2E_{t,b}}{M_h}$, $y_{t,b} = 1- x_{t,b}$,  $\kappa_W = \frac{m_W^2}{M_h^2}$, $\kappa_t = \frac{m_t^2}{M_h^2}$, and $\gamma_t = \frac{\Gamma_t^2}{M_h^2}$. Using these definitions, $s_{1,2}$ can be written as:
\small
\begin{eqnarray}
s_1 &=& -P^2 - p_1^2 + 2 P\cdot p_1 = M_h^2 + m_b^2 - 2 M_h E_b \approx M_h^2 \left(1 - \frac{2 E_b}{M_h} \right)  \equiv M_h^2 y_b, \nonumber \\
s_2 &=& -P^2 - p_2^2 + 2 P\cdot p_2 = M_h^2 + m_t^2 - 2 M_h E_t = M_h^2 \left(1 - \frac{2 E_t}{M_h} + \frac{m_t^2}{M_h^2}\right) \equiv M_h^2 y_t + m_t^2 \label{s1s1},\\\nonumber
\end{eqnarray}
where $m_b$ is neglected.
With $ds_1 = -2 M_h d E_b$ and $ds_2 = -2 M_h d E_t$, the differential decay width can be written as
\small
\begin{eqnarray}
d\Gamma &=& 
 \frac{N_c}{(2\pi)^3} \frac{1}{8 M_h} \langle |T|^2 \rangle \; dE_b dE_t 
= \frac{N_cG_F^2 }{64 \pi^3 }  m_t^2 M_h^3 \frac{\Gamma_0}{ y_t^2 + \gamma_t \kappa_t}  \; dx_b dx_t ,
\end{eqnarray}
where $G_F = g /(4 v M_W)$. This result agrees with the formula in Ref.~\cite{Djouadi}.



\end{document}